\documentclass[runningheads]{llncs}
\usepackage[T1]{fontenc}
\usepackage{adjustbox}
\usepackage{rotating}
\usepackage{array}
\usepackage{graphicx}
\usepackage{subfig}    
\usepackage{rotating}   
\usepackage{amsmath,amssymb,amsfonts}

\usepackage{amsthm}
\usepackage{mathtools}
\usepackage{textcomp}
\usepackage{xcolor}
\usepackage{subcaption}
\usepackage{algorithm, algpseudocode}
\mathtoolsset{showonlyrefs=true}
\usepackage{enumitem}
\usepackage[hidelinks]{hyperref}

\begin{document}
\thispagestyle{empty}
\onecolumn   
Accepted to: the 18th International Symposium on Distributed Autonomous Robotic Systems (DARS 2026).

Copyright may be transferred without notice, after which this version may no longer be accessible.
\newpage

\title{Safe and Energy-Aware Decentralized PDE-Constrained Optimization-Based Control of Multi-UAVs for Persistent Wildfire Suppression}

\titlerunning{Multi-UAVs for Persistent Wildfire Suppression}
\author{Longchen Niu\inst{1}\orcidID{0009-0001-1901-7991} \and
Gennaro Notomista\inst{1}\orcidID{0000-0002-1478-2790} }
\authorrunning{L. Niu, and G. Notomista}
%
\institute{University of Waterloo, Waterloo, Ontario, Canada\\
\email{l3niu@uwaterloo.ca, gennaro.notomista@uwaterloo.ca}}
\maketitle              
\begin{abstract}
This paper presents a safe and energy-aware optimization-based control framework for multi-UAV wildfire suppression under localization and motion uncertainties. We first develop a centralized density-based controller that couples UAV motion and water deployment in a wildfire-specific control Lyapunov function. This framework is then extended to a decentralized setting suitable for large-scale operations using only local information. The controllers use control barrier function constraints to enforce both danger zone avoidance and the ability to reach a charging region. Simulations and real quadcopter experiments demonstrate the controller's effectiveness in fire suppression while preserving safety and energy sufficiency over multiple charge cycles. 

\end{abstract}

\section{Introduction}
Wildfires, caused by both human activity and climate change, are increasing in frequency and severity, creating growing risks to ecosystems, infrastructure, human safety, and global climate \cite{fire_impact_onclimate,fire_by_man,fire_impactedbyclimatechange}. Since fires become increasingly difficult to contain as they grow, rapid detection and early suppression are critical to eliminate damage to the environment and prevent small sparks from escalating into catastrophic disasters. Autonomous aerial systems, specifically Unmanned Aerial Vehicles (UAVs), offer a promising solution because they can access dangerous or remote regions, operate without risks to human pilots, and coordinate as multi-UAV teams for large-scale monitoring and suppression tasks. 

Many existing UAV-based wildfire systems focus only on sensing, coverage, and monitoring as early fire detection alarms. More recent works have studied the UAV's dynamics during heavy suppressant deployment and optimal control strategies for protecting critical regions. 
Studies with long-range, high-payload fixed-wing UAVs have also been conducted through heuristic-based algorithms for wildfire monitoring and suppression. While these UAVs are suitable for large-area missions on the scale of provinces, lower-cost quadcopter teams can be deployed more flexibly in remote areas on mobile stations, for precise deployment near fire fronts. However, quadcopter-based applications also raise the challenges of limited onboard resources, safety in proximity to burning areas, and persistent recharging abilities. 
While existing approaches provide promising frameworks for wildfire response, they often do not address decentralized long-term operations with persistent energy feasibility and formal safety guarantees required for fully autonomous systems. Specifically, rapid wildfire response requires quadcopters to react to an evolving and uncertain fire field while avoiding unsafe regions, balancing limited onboard resources, and maintaining the ability to return to charging stations at all times. 

In this paper, we formulate wildfire suppression as an optimization-based control problem for multi-UAV systems, which takes into account localization and motion uncertainties. We first extend our previous centralized safe and energy-aware controller to wildfire suppression tasks by designing a Control Lyapunov Function (CLF) that couples UAV motion and water deployment within a unified density-based Partial Differential Equation (PDE) framework \cite{ITSC_Energy,ITSC_Energy_arxiv}. We then decentralize this framework for large-scale operations while enforcing spatial safety and energy sufficiency constraints with formal guarantees through Control Barrier Functions (CBFs). Finally, the proposed framework is validated in both simulations and quadcopter experiments to demonstrate the effectiveness of the controller for long-term wildfire monitoring and suppression operations.

\section{Related Work and Contributions}
\label{sec: related work}
With the increase in wildfire incidents around the globe due to climate change, wildfire suppression has been studied in response \cite{fire_impactedbyclimatechange}. In particular, the work in \cite{fire_study_no_tech} highlights the importance of early response since containment becomes more difficult as fires grow. This motivates the studies on autonomous detection and suppression systems that can reduce fire growth and eliminate the risk in the early stages. 
From a formal mathematical perspective, the authors of \cite{Fire_PDE_Boundary} formulated wildfire mitigation as a PDE-based control problem where temperature regulation is achieved via boundary control. While providing a formal model for the temperature field that is employed in this work, control of a fixed boundary can be impractical to implement due to its immobility compared to UAVs.

A large body of work has studied sensing, coverage, and information gathering for distributed robotic systems that can be adapted for early fire detection. Decentralized coverage-control methods have been used for information gathering without active communication during operation, as seen in \cite{gan2014online,8460846,doi:10.1155/2015/286080}. Other approaches share limited information online, such as scalar measurements for gas-leak localization \cite{NANAVATI2024102503}. Continuous field estimation methods suitable for wildfire models have also been proposed, such as the fixed Gaussian base method in \cite{7365431}. However, this representation can require an unnecessarily large number of robots to accurately approximate the environment \cite{7989245}. More flexible Gaussian-process-based approach using online distributed consensus learning is studied in \cite{luo2019distributed}, with recent extensions to goal-directed planning in \cite{11128099}. Nevertheless, these methods still rely on strong communication assumptions, such as persistent Voronoi-neighbor connectivity, and consensus steps can temporarily stall the team, delaying rapid response to fire. 

Within wildfire-specific UAV monitoring studies, the authors of \cite{11005691} formulate an energy-efficient multi-UAV coverage problem where camera-equipped UAVs track an evolving fire perimeter while considering dynamics, battery limits, charging stations, collision avoidance, and communication constraints. However, the framework is for monitoring only without active suppression, and safety is embedded through rewards and penalties rather than formal invariance guarantees seen in our framework.
Detailed studies on vehicle dynamics during the suppression phase have been conducted in \cite{SIngleDropStudy}, which proposes an optimal-control approach for a single UAV to drop a large suppressant payload near a fire by exploiting the vehicle's time-varying mass after release. While this enables accurate payload delivery with reduced heat exposure, it focuses on a single aggressive drop maneuver rather than long-horizon multi-UAV coordination. 

More similar to our objectives, the authors of \cite{Fire_PDE_OC} formulate aerial wildfire mitigation as a PDE-constrained optimal control problem with a UAV and water-flow control. This method optimizes the vehicle trajectory and watering strategy to protect an exclusion zone. However, this centralized optimal control approach lacks the rapid response time to unmodeled disturbances, which our proposed optimization-based control excels at. Although the authors stated that a multi-robot extension is possible, it was not studied explicitly. The formulation also assumes unlimited battery resources, whereas our framework explicitly enforces energy sufficiency with a multi-UAV team.
Moreover, our framework also accounts for localization and motion noises common in practice, while providing safety guarantees for the entire team. 
The work in \cite{fire_detect_large} studied large-scale wildfire detection using decentralized force-based UAV swarms, and extends to wildfire suppression in \cite{fire_supp_large}. These methods use dynamic partitioning, force-based interactions, and local coordination rules, making them scalable but heavily heuristic-dependent. More importantly, the studies consider large fixed-wing aircraft with a travel range of \(1000\)~km, so energy sufficiency is not explicitly studied. While this is reasonable for long-range aircraft, many practical wildfire scenarios could benefit from lower-cost, multi-UAV quadcopter swarms that can be deployed more flexibly in remote areas with mobile stations, where energy limitations become a central constraint. In contrast, our work formulates wildfire suppression as a decentralized optimization-based control problem, where CBF constraints formally enforce spatial and energy safety, with quantifiable performance through the CLF constraint, for persistent wildfire suppression. 

To address these gaps in the existing literature, we build on the safety- and energy-aware techniques introduced in our previous works \cite{ITSC_Energy,MRS_paper,ais_paper} and make the following contributions:

\begin{enumerate}[label=(\roman*)]
   \item We leverage techniques developed for the safety and energy constraints in our previous works to formulate centralized density control for wildfire suppression as a new problem governed by a coupled system of PDEs. Within this framework, we design a centralized controller using a novel Lyapunov function that jointly captures UAV motion and water deployment while accounting for motion and localization uncertainties.
    
    \item Unlike our previous energy-aware formulation, we develop a decentralized constrained kinodynamic motion model for path planning. This model enables each robot to independently plan a safe path to a charging station and sustain operation over multiple charging cycles.

    \item  We integrate the coupled wildfire-suppression objective, decentralized kinodynamic planning, and CBF safety constraints into a decentralized optimization-based controller for wildfire-suppression tasks that formally guarantees safety.
    
    \item We demonstrate the effectiveness of the proposed controllers in simulations and with a real multi-UAV system. 
\end{enumerate}

\section{Problem Formulation}
In this work, we consider a multi-UAV wildfire mitigation problem over a bounded domain \(\Omega \subset \mathbb{R}^2\), where wildfire evolves as a continuous temperature--fuel field and a team of $N$ suppressant-carrying quadcopters is deployed to contain its spread. Unlike protected-region formulations, our objective is to limit the spatial growth of the fire region itself. Specifically, the swarm should move toward the boundary of each active fire, avoid high-temperature fire regions, and maintain safe battery margins for return to charging stations.

\subsection{Fire Dynamics Model}
Let $T(r,t)$ denote the temperature field and $S(r,t)$ denote the remaining fuel mass at location $r \in \Omega$ and time $t \ge 0$. Motivated by the wildfire models in \cite{Fire_PDE_Boundary}, we consider the temperature-based fire dynamics
\begin{equation}
\begin{cases}
    \frac{\partial T}{\partial t} = \eta \Delta T - v_f \nabla T + A\big(Sr(T)-C_c(T-T_a)\big) + u_s(r,t),\\
    \frac{\partial S}{\partial t} = - C_s S r(T),
\end{cases}
\label{eq:fire_dynamics}
\end{equation}
where $\eta > 0$ is the thermal diffusivity, \(v_f: \Omega \times [0,+\infty) \to \mathbb{R}^2\) is the wind field, \(C_c\) is the cooling coefficient to the ambient temperature \(T_a\), and $S r(T)$ represents the rate at which fuel is consumed due to burning, with the Arrhenius law:
$r(T)=\exp\!\left(-\dfrac{\phi}{T-T_a}\right)$ for $T>T_a$, and $r(T)=0$ for $T\leq T_a$, with $\phi > 0$. The coefficients \(A\) and \(C_s\) characterize the maximum temperature increase and fuel consumption rates, with $u_s(r,t)$ as the suppression input. Following \cite{Fire_PDE_OC}, the UAV's suppression output is modeled as:
\begin{equation}
u_s(r,t)=\sum_{i=1}^N - f_i(t)K_{\mathrm{spray}}(r,x_i),
\label{eq:suppression flow}
\end{equation}
where \(K_{\mathrm{spray}}(r,x_i) =e^{-\frac{1}{2\sigma_s^2}\|r-x_i\|^2}\) is a Gaussian shaping function, characterized by \(\sigma_s\), representing the suppressant spray shape at $x_i(t)$, the position of robot $i$, multiplied by the positive controlled water flow rate \(f_i(t)\).

\subsection{Robot Density Model}
Following the seminal work in \cite{ANNUNZIATO2013487,Archer2004DynamicalDF}
and our recent works in \cite{ITSC_Energy,RalPaper,MRS_paper,ais_paper}, the spatial density of a stochastic multi-robot system can be represented by the Probability Density Function (PDF): \(\rho(r, t) = \sum_{i=1}^{N} \rho_i\), where each \(\rho_i\) captures the belief-weighted physical probability mass of robot \(i\) under localization uncertainty. The collective density model of the system evolves according to the Fokker-Planck equation:
\begin{equation}
    \frac{\partial \rho(r, t, u)}{\partial t} = \sum_{i \in N} \bigg[-\nabla_i(u_i \  \rho_i(r,t)) + D_\rho \Delta_i \rho_i(r, t) \bigg],
    \label{eq:FP_equation}
\end{equation}
which provides a deterministic macroscopic model of the system density despite the presence of uncertainties. In particular, localization noise is captured through the precision matrix \(\Sigma\) in the Gaussian PDF \(\rho_i (r,t) = \frac{1}{\sigma \sqrt{2\pi}} \exp\left(-\frac{1}{2} \Sigma (r-x_i)^2\right)\) at the measured position \(x_i\), while the motion noise is incorporated through the diffusion coefficient \(D_\rho\) in \eqref{eq:FP_equation}. As a result, instead of tracking individual robots through separate stochastic differential equations, equation \eqref{eq:FP_equation} governs the deterministic evolution of the entire team's probabilistic spatial distribution under the control input \(u=[u_1^T,\dots,u_N^T]^T\), where \(u_i\in\mathbb{R}^2\) is the velocity command of robot \(i\) bounded by the maximum velocity \(u_{\max}\) in \(\mathcal{U}:=\{v\in\mathbb{R}^2:\|v\|\leq u_{\max}\}\).
This density framework serves as the fundamental system dynamics model for the remainder of this paper. 

\section{Controller Design}
In this section, we first present the centralized controller from our previous work \cite{ITSC_Energy} and the proposed centralized wildfire suppression controller with the novel CLF, followed by the decentralized controller from \cite{MRS_paper} and the proposed decentralized wildfire suppression controller. 

\subsection{Safe and Energy-aware Centralized Controller}
In \cite{ITSC_Energy}, we developed a centralized PDE-based controller for smart transportation applications with kinodynamic path planning. This controller serves as the foundation for the present work and is given by:
\begin{equation}
    \begin{aligned}
    \min_{u_i \in \mathcal{U},\,s} &\quad \|u\|^2 + \gamma s\\
    \quad \text{s.t.} &\quad \alpha_v V(\rho) + \dot{V}(\rho,u) - s\leq 0\\
    &\quad \alpha_h h_s(\rho) + \dot h_s(\rho,u) \geq 0\\
    &\quad \alpha_E h_{E_i}(x_i) + \dot h_{E_i}(x_i, u_i) \geq 0 \quad \forall i \in N,
\end{aligned}
\label{eq: ITSC model}
\end{equation}
with positive coefficients \(\alpha_v, \alpha_h\), and \(\alpha_E\). The control objective is to minimize the control effort with a non-negative slack variable \(s\), scaled by \(\gamma\), to relax the CLF constraint when needed. This CLF guides the team of robots towards a predefined target PDF \(\rho_d\), defined as: \(V(\rho) = \int_\Omega (\rho_d - \rho)^2 \ dr, \dot V(\rho, u) = -2 \int_\Omega (\rho_d - \rho) \dot \rho \ dr,\)
where the time evolution \(\dot \rho\) follows the Fokker-Planck equation \eqref{eq:FP_equation}. The second constraint in \eqref{eq: ITSC model} enforces spatial safety via the CBF: \(h_s(\rho) = \epsilon - \int_\mathcal{A} \rho^2 \ dr, \,  \dot h_s(\rho, u) = -2 \int_\mathcal{A} \rho \ \dot \rho \ dr, \)
where \(\epsilon\) bounds the allowed residual belief mass due to localization noise within the unsafe region \(\mathcal{A} \subset \Omega\). As shown in \cite{ais_paper}, this CBF constraint guarantees forward invariance of the density set \(\{ \rho: h_s(\rho) \geq 0 \}\). 
Finally, the energy sufficiency of the team is enforced with the per-robot CBF and its derivative: 
\begin{equation}
\begin{aligned}
    &h_{E_i}(x_i) = E_i - E_{\min} - P(g_i,x_i), \quad \dot h_{E_i}(x_i, u_i) = -c_1 \|u_i\|^2 - c_2 - \dot P(g_i,x_i),
\end{aligned}
\label{eq: energy detail def}
\end{equation}
where \(E_i \in [0,1]\) is robot \(i\)'s remaining energy, \(E_{\min}\) is the minimum safe energy level, and the terms \(c_1 \|u_i\|^2\) and \(c_2\) encode motion-induced and constant operation energy cost, respectively. The value \(P(g_i, x_i)\) is the required energy for robot \(i\), at position \(x_i\), to reach a charger region \(\mathcal{C} \subset \Omega\) following the path \(g_i:[0,t_f]\rightarrow\Omega\). The rate of change of this path-based energy-to-charge value is denoted as \(\dot P(g_i,x_i)\). In \eqref{eq: ITSC model}, \(g_i\) is generated using the following CBF-constrained kinodynamic model:    
\begin{equation}
\begin{aligned}
    &\dot x_i (t) = u_i(t)\\
    \text{s.t.} \ & \alpha_s h_s(\rho(t)) + \dot h_s (\rho(t), u(t)) \geq 0 \\
    & x_i(t_f) \in \mathcal{C},
\end{aligned}
\label{eq: pathplanning model}
\end{equation}
The model is implemented using an RRT with constrained state-based steering, producing \(g_i\) as a sequence of feasible waypoints \cite{rrt_paper}. Other robots are treated as stationary during planning and are jointly coordinated during controller execution to maintain safety.
Importantly, \cite{ITSC_Energy_arxiv} establishes the feasibility of \eqref{eq: ITSC model} for safe and energy-aware centralized control. 

\subsection{Proposed Centralized Wildfire Suppression Controller}
Building on the foundations of \eqref{eq: ITSC model}, we extend the framework for the wildfire application with the temperature model \eqref{eq:fire_dynamics}. 

To define the task objective, we first introduce the suppression deployment region: \(\mathcal{D}(t) := \{r \in \Omega : T(r,t) \in [T_{\mathrm{low}}, T_{\mathrm{high}}]\},\)
where two environment-specific parameters, \(T_{\mathrm{low}}, T_{\mathrm{high}}\), define the band where suppressant deployment is desired, typically near the boundary of active fire. The temperature field \(T(r,t)\) is then scaled up within \(\mathcal{D}(t)\) to construct a global target PDF \(\rho_d(r,t)\) over \(\Omega\), representing the desired spatial concentration of water for suppression.

Then, to capture the behaviors of UAVs with onboard water reservoirs, define:
\begin{equation}
    \rho_w(r,t) = \sum_{i=1}^N w_i(t)\,(K_{\mathrm{spray}} * \rho_i)(r,t) = \sum_{i=1}^N w_i(t)\int_{\Omega} K_{\mathrm{spray}}(r,\xi)\,\rho_i(\xi,t)\,d\xi,
\end{equation}
as the spatial distribution of water deployment, obtained by convolution of the suppressant spray kernel, \(K_\mathrm{spray}\) in \eqref{eq:suppression flow}, with each robot’s normalized localization density \(\rho_i(r,t)\), and weighted by the onboard water level \(w_i(t)\).

Therefore, we can define the task Lyapunov function and its derivative as:
\begin{equation}
\begin{aligned}
V(\rho_w,t) &= \frac{1}{2}\int_{\Omega} \big(\rho_d(r,t) - \rho_w(r,t)\big)^2\,dr, \\
\dot V(\rho_w,t, u, f) &= \sum_{i=1}^N \int_{\Omega} \bigg(\rho_d(r,t) - \rho_w(r,t)\bigg) \\& \quad \bigg(- w_i(t) K_{\mathrm{spray}} *\dot \rho_i(r,t, u_i) + f_i(t) K_{\mathrm{spray}} *\rho_i(r,t) \bigg) \,dr,
\end{aligned}
\label{eq:clf}
\end{equation}
where \(f = [f_1,\dots,f_N] = [-\dot w_1, \dots, -\dot w_N]\) is the controlled water flow rate in \eqref{eq:suppression flow}. With this formulation, the CLF drives the team to both transport and deploy water towards the target density \(\rho_d\) concentrated over the deployment zone \(\mathcal{D}\).

Next, the unsafe region used in the spatial safety CBF is defined as:
\(\mathcal{A}(t) := \{r \in \Omega : T(r,t)\ge T_{\mathrm{high}}\},\)
which is treated as a no-fly zone to avoid thermal damage to the UAVs. 
The resulting centralized wildfire suppression controller is: 
\begin{equation}
    \begin{aligned}
    \min_{u, \, f,\,s} &\quad \|u\|^2 + \zeta \|f\|^2 + \gamma s\\
    \quad \text{s.t.} &\quad \alpha_v V(\rho_w) + \dot{V}(\rho_w, u, f) - s\leq 0\\
    &\quad \alpha_h h_s(\rho) + \dot h_s(\rho,u) \geq 0\\
    &\quad \alpha_E h_{E_i}(x_i) + \dot h_{E_i}(x_i, u_i) \geq 0 \quad \forall i \in N,
    \end{aligned}
    \label{eq: centralized controller}
\end{equation}
where \(\zeta\) balances the cost of water transportation and deployment, while the energy CBF remains unchanged. The optimization problem \eqref{eq: centralized controller} yields a density-based swarm controller that drives the robots toward \(\mathcal{D}\) to deploy water based on \(\rho_d(r,t)\), avoids the unsafe area \(\mathcal{A}\), and preserves sufficient battery level for long-term operations. Because the CLF is relaxed by \(s\), feasibility is determined by the hard CBF constraints. The spatial CBF is feasible under the bound established in \cite{ais_paper}, while \cite{ITSC_Energy_arxiv} shows that the energy CBF has the same admissible control set. Hence, controller \eqref{eq: centralized controller} remains feasible under the same bound shown in \cite{ais_paper}.

\begin{remark}
In this work, we set \(\dot{\rho}_d=0\) because the future evolution of the target is unknown. An alternative is to separate transportation and suppression using \(V=\frac{1}{2}\|\rho_d-\rho_w\|^2+\frac{1}{2}\|T-T_d\|^2\), with the fire dynamics \eqref{eq:fire_dynamics} entering through both \(\dot T\) and \(\dot\rho_d\). Like the proposed formulation, this design permits deployment while the UAVs travel, but requires a known fire model and additional computation. Furthermore, the resulting \(\dot V\) differs only in the factor multiplying \(f_iK_{\mathrm{spray}} *\rho_i\): it uses the temperature error rather than the reservoir-weighted density error.

In this work, we instead employ a single \(\rho_d(r,t)\), constructed from the scaled temperature field \(T(r,t)\). This reduces the computational cost of the controller while allowing the relative priority of transportation and deployment to be tuned through \(\zeta\). Moreover, allowing water deployment during transportation can pre-cool areas surrounding the active fire, reducing their risk of ignition, while decreasing the UAVs' payload mass and associated energy consumption.
\end{remark}

\subsection{Decentralized Safe Controller}
In large-scale applications, the centralized requirement of real-time communication with the entire team is often impractical. To address this, decentralized controllers were developed, where each robot operates with only locally available information within a detection radius \(d\). In our work \cite{MRS_paper}, we developed a decentralized safe density controller of the form: 
\begin{equation}
    \begin{aligned}
    \min_{u_i\in \mathcal{U},\,s} &\quad \|u_i\|^2 + \gamma s\\
     \quad \text{s.t.} &\quad \alpha_v{V}_i + \dot{V}_{i} - s\leq 0, \quad \alpha_h {h}_{s,i} + \dot{h}_{s,i} \geq \delta_i,
    \end{aligned}
    \label{eq: MRS Model}
\end{equation}
where the Lyapunov and safety barrier functions are decentralized over the local neighborhood set defined by a distance-based communication graph, \(\Delta\)-disk graph \cite{mesbahi2010graph}, as:
\(\mathcal{N}_i = \left\{ j \in N\, \middle| \, \| {x}_j - {x}_i \| \leq d \right\}\). 

Importantly, the relaxation term \(\delta_i\) is computed using the worst-case neighbor motion prediction, \(\dot \rho_{\mathcal{N}_i}^\mathrm{worst}\), to ensure safety even when all neighbors of robot \(i\) act aggressively. However, unlike the centralized controllers discussed above, \eqref{eq: MRS Model} does not account for robot energy sufficiency or suppressant deployment tasks. 

\subsection{Proposed Decentralized Wildfire Suppression Controller}
\label{subsec: decentralized controller}
To extend the decentralized controller to wildfire suppression tasks, we first localize the CLF \eqref{eq:clf} using the neighborhood definition above. Specifically, the global density \(\rho(r,t)\) is replaced by the local detectable density \(\rho_{\mathcal{N}_i}(r,t)\) as the sum of the densities of all robots in \(\mathcal{N}_i\). 
Unlike \cite{MRS_paper}, where the neighbors are considered stationary for the CLF, we assume the neighbors maintain their last commanded motion, due to the dominant effects of inertia in drone dynamics, with their water flow rates, \(f_j\)'s, set to zero as a conservative worst-case choice.     

Next, we develop decentralized energy awareness for this application. In \eqref{eq: ITSC model} and \eqref{eq: centralized controller}, energy safety was enforced via a CBF-constrained kinodynamic model \eqref{eq: pathplanning model}. In the decentralized case, however, neighboring robots cannot solve a shared optimization problem online, as it will result in a more computationally demanding distributed controller \cite{distributed_opt_survey}. 
Therefore, the path planner replaces the global spatial CBF \(h_s\) in \eqref{eq: pathplanning model} with the local CBF constraint used in \eqref{eq: MRS Model}. In particular, the initial planning step enforces \(\alpha_h h_{s,i}+\dot h_{s,i}\geq\delta_i\). Neighboring robots are subsequently represented by their predicted initial velocities \(u_j(0)\) based on the worst-case prediction, avoiding a shared online optimization. Because the path-based energy CBF is constructed consistently with the local safety constraint, the two CBFs share the same admissible control set. Thus, the feasibility conditions established for \eqref{eq: MRS Model} apply directly to the decentralized wildfire-suppression controller:
\begin{equation}
    \begin{aligned}
    \min_{u_i,f_i,s} &\quad \|u_i\|^2 + \zeta f_i^2 + \gamma s\\
     \quad \text{s.t.} &\quad \alpha_v{V}_i (\rho_{\mathcal{N}_i}, w_{\mathcal{N}_i}) + \dot{V}_{i}(\rho_{\mathcal{N}_i}, u_i, w_i, f_i) - s\leq 0\\
    &\quad \alpha_h {h}_{s,i}(\rho_{\mathcal{N}_i}) + \dot{h}_{s,i}(\rho_{\mathcal{N}_i}, u_i) \geq \delta_i\\
    &\quad \alpha_E h_{E_i}(x_i) + \dot h_{E_i}(x_i, u_i) \geq 0.
    \end{aligned}
    \label{eq: decentralized model}
\end{equation}

As discussed in Sec.~\ref{sec: related work}, several studies have considered online sensing and data sharing \cite{11128099,luo2019distributed,NANAVATI2024102503}, but none are directly applicable to the wildfire suppression setting. We therefore adopt a simple decentralized sensing model: each robot maintains a local temperature map within its detection radius, and the certainty of this map decays linearly over time as the environment is constantly evolving. This map is shared immediately with neighbors, but only through one-hop communication: maps received from neighbors are not shared. 

\begin{remark}
    Unlike the locally shared sensing map, the unsafe region \(\mathcal{A}\) must be broadcast globally to ensure a feasible path to the charger region \(\mathcal{C}\) can be found.
\end{remark}

For numerical implementation of the controllers \eqref{eq: centralized controller} and \eqref{eq: decentralized model}, the continuous PDEs are spatially discretized using the finite-difference method on a uniform two-dimensional grid, yielding a finite-dimensional system of ODEs. This ODE formulation does not replace the PDE systems, as they are essential to describe the evolution of the spatial UAVs and water densities while incorporating uncertainty and motion noise, but rather provides a standard computational approximation. The adopted discretization is well suited to the smooth Gaussian density functions considered in this work and is commonly used for implementing PDE-based continuous-time control systems. Further details regarding the discretization scheme are provided in our previous work \cite{ais_paper}.

This completes the controller design for both the centralized and decentralized wildfire suppression formulations. In the following sections, we evaluate the proposed controllers in both simulations and experiments.

\section{Simulation Results}
To evaluate the effectiveness of the proposed controllers, this section presents the simulation setup, performance metrics, and comparative results for centralized and decentralized fire-suppression controllers.

The simulations were conducted in a square environment with a side length of \(4\)~m and a spatial resolution of \(0.1\)~m. Three initial fires, centered at \((-0.3,0.4),\) \((-0.1,0.0),(0.5,-0.4)\), ignite the forest at \(t = 0\)~s. Each simulation runs for \(50\)~s with a timestep of \(0.05\)~s. In the uncontrolled baseline, the fire extinguishes naturally after approximately \(45\)~s once the fuel is depleted, corresponding to the complete burn-down of the forest. The charger region \(\mathcal{C}\) is a circle of radius \(0.5\)~m centered at \((-1.2, -1.2)\). The field size, UAV speed, battery depletion, and charging rates are chosen so that energy feasibility becomes a critical concern within the simulation and experiment time frame. Therefore, the results should be analyzed as the relative time scales between the fire dynamics, UAV motion, suppressant effects, and energy parameters, rather than discrete real-time durations. Longer missions are directly implementable by rescaling the parameters. 

The battery level \(E_i(t)\) and spatial safety measurement \(h(\rho)\) are used to evaluate the survivability of the proposed controllers. To quantify fire suppression performance, we define an ignition temperature \(T_{\mathrm{ignite}}\), since the Arrhenius law used in \eqref{eq:fire_dynamics} does not define an explicit ignition threshold. The active-fire region at time \(t\) is defined as
\(
\mathcal{F}(t) := \left\{ r \in \Omega \;:\; T(r,t) \geq T_{\mathrm{ignite}} \right\},
\)
with the corresponding fire area 
\(
A_f(t) = |\mathcal{F}(t)| = \int_{\Omega} \mathbf{1}_{\{T(r,t)\geq T_{\mathrm{ignite}}\}}\,dr.
\)
The multi-UAV team should reduce the growth of \(A_f(t)\) and, ideally, drive it to zero before the \(45\)~s natural burn-down mark, indicating active extinction by water deployment. 

Since the underlying model is stochastic, the following results are obtained from \(100\) simulations with different noise realizations. At each timestep, \(x_i^\mathrm{measured}\) is sampled from a Gaussian distribution centered at \(x_i^\mathrm{true}\) with \(\Sigma\); the control \(u_i^*\) is then computed with \(x_i^\mathrm{measured}\) using \eqref{eq: centralized controller} or \eqref{eq: decentralized model}, and \(x_i^\mathrm{true}\) is updated with Gaussian motion noise defined by \(D_\rho\) in \eqref{eq:FP_equation}.
Simulations were conducted on a MacBook Pro with an Apple M4 Pro chip (10-core CPU) and \(24\)~GB of unified memory using the MOSEK Fusion API for Python \cite{mosek}. Code is available at: \href{https://github.com/erablab/PDE_based_Wildfire_Suppresion_Controllers}{\nolinkurl{github.com/erablab/PDE_based_Wildfire_Suppresion_Controllers}}.

\begin{figure*}[t]
\centering
    \setlength{\tabcolsep}{2pt}

    \begin{minipage}[t]{0.51\textwidth}
        \vspace{0pt}
        \centering
        \includegraphics[width=\linewidth]{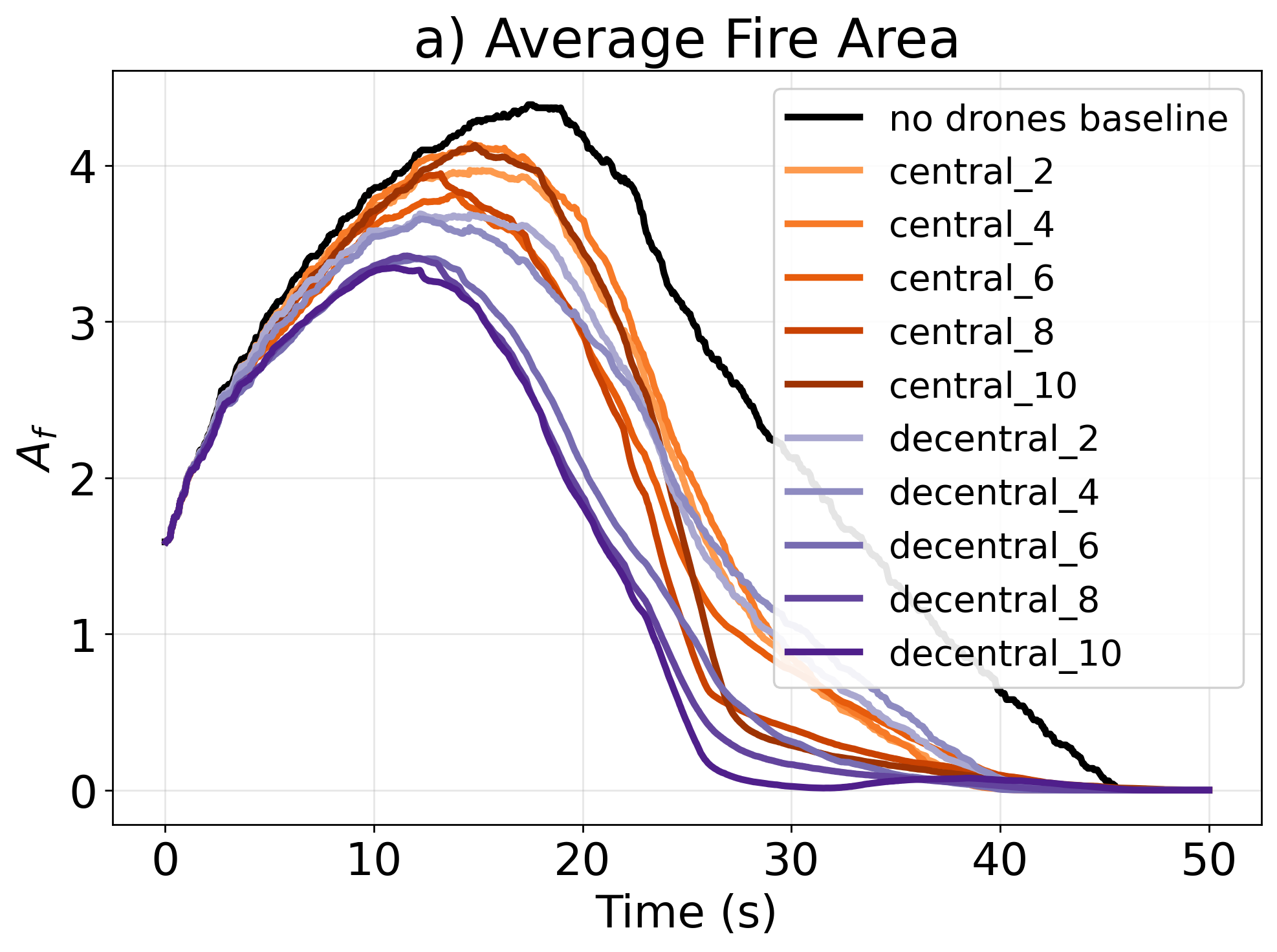}
    \end{minipage}
    \hfill
    \begin{minipage}[t]{0.48\textwidth}
        \vspace{0.4em}
        \centering

        \includegraphics[width=0.48\linewidth]{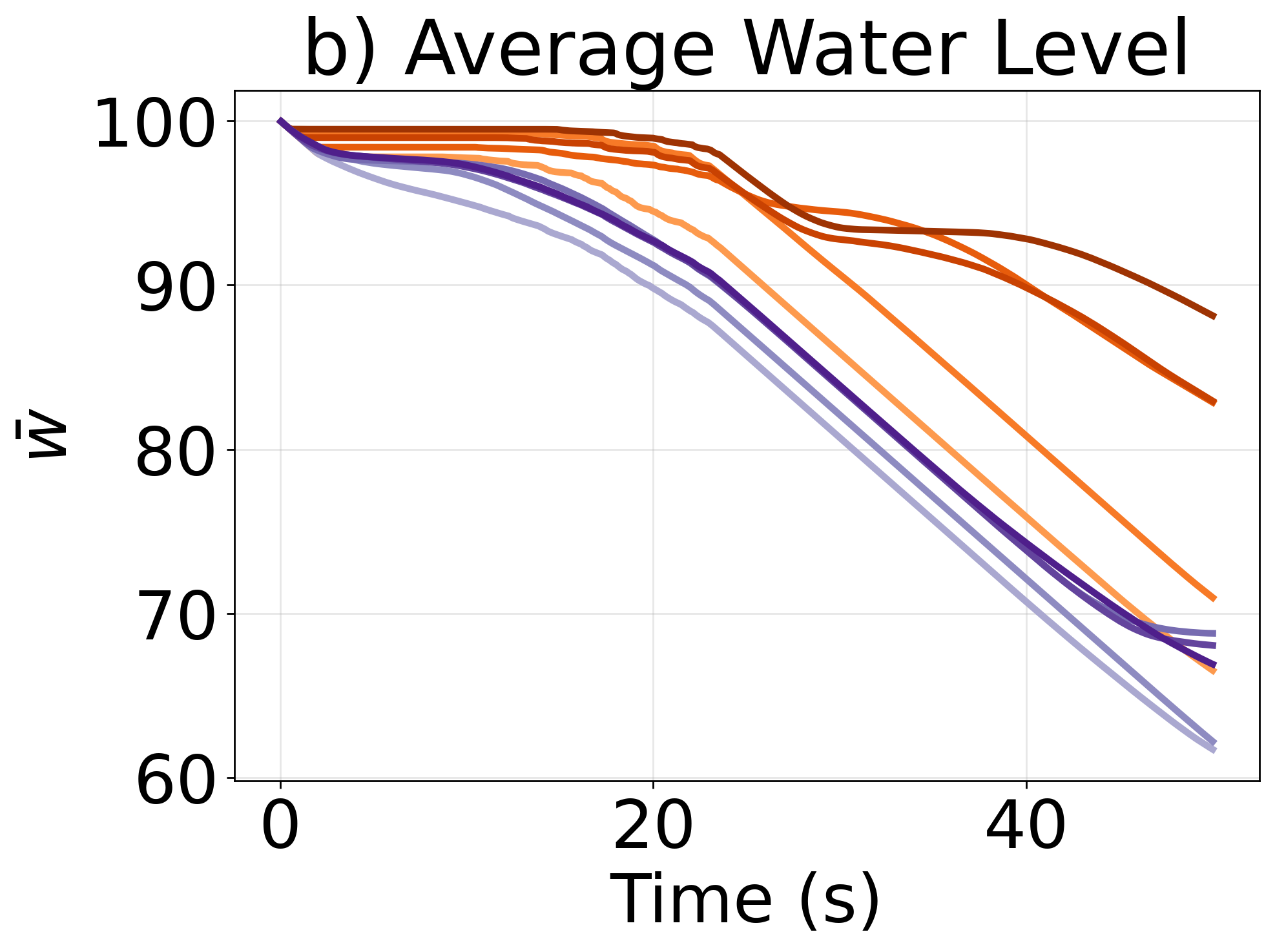}
        \hfill
        \includegraphics[width=0.48\linewidth]{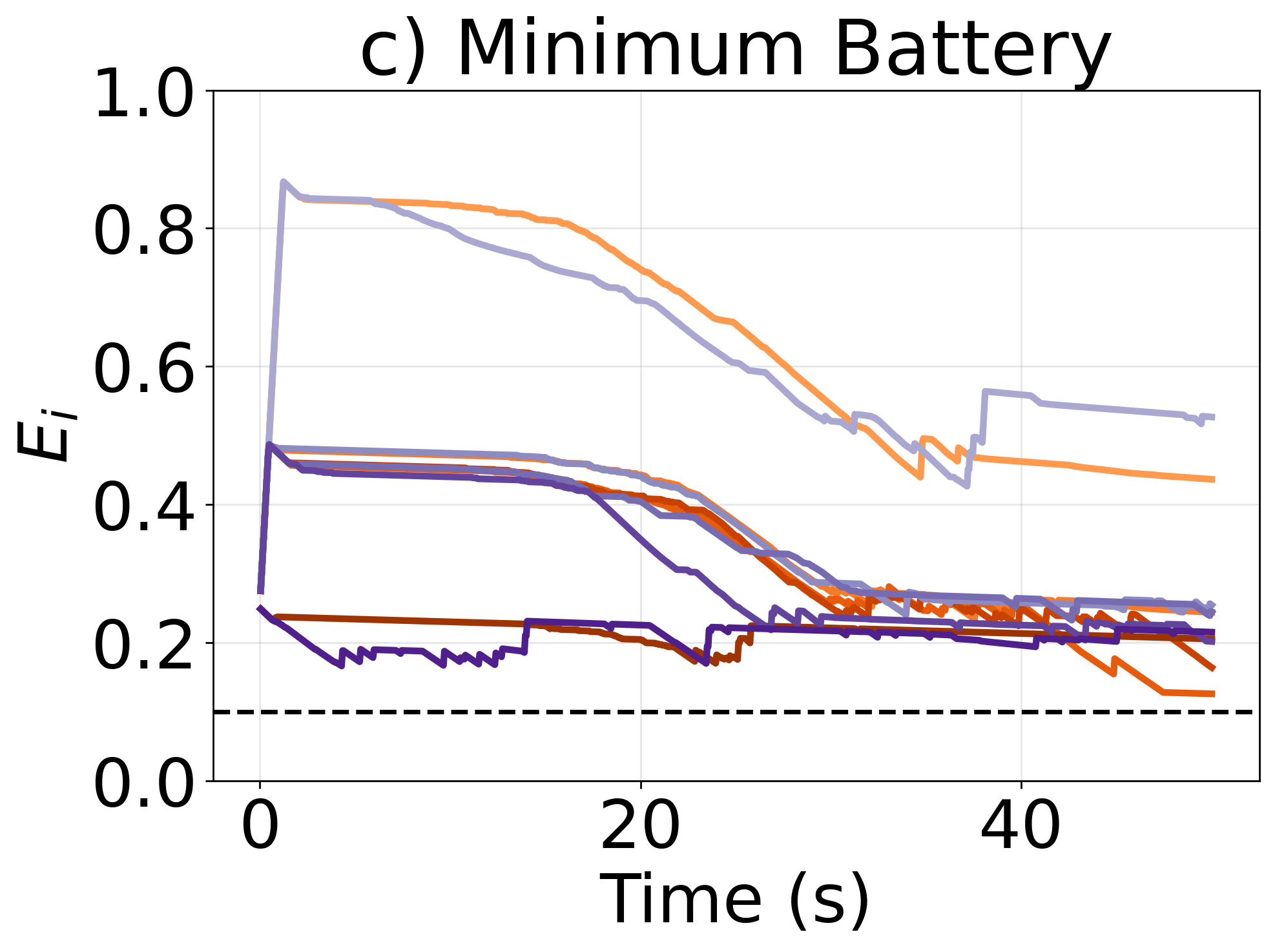}


        \includegraphics[width=0.48\linewidth]{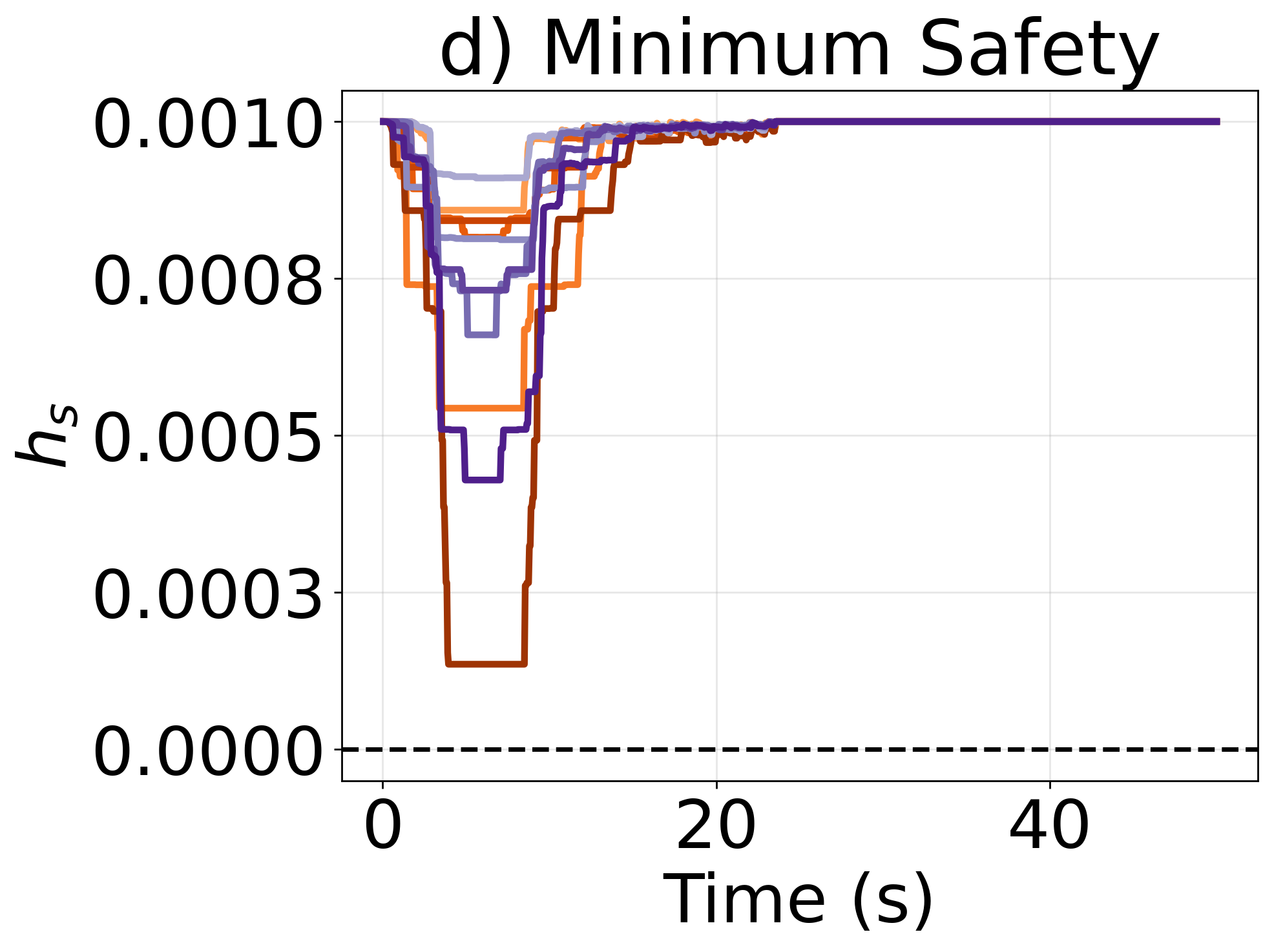}
        \hfill
        \includegraphics[width=0.48\linewidth]{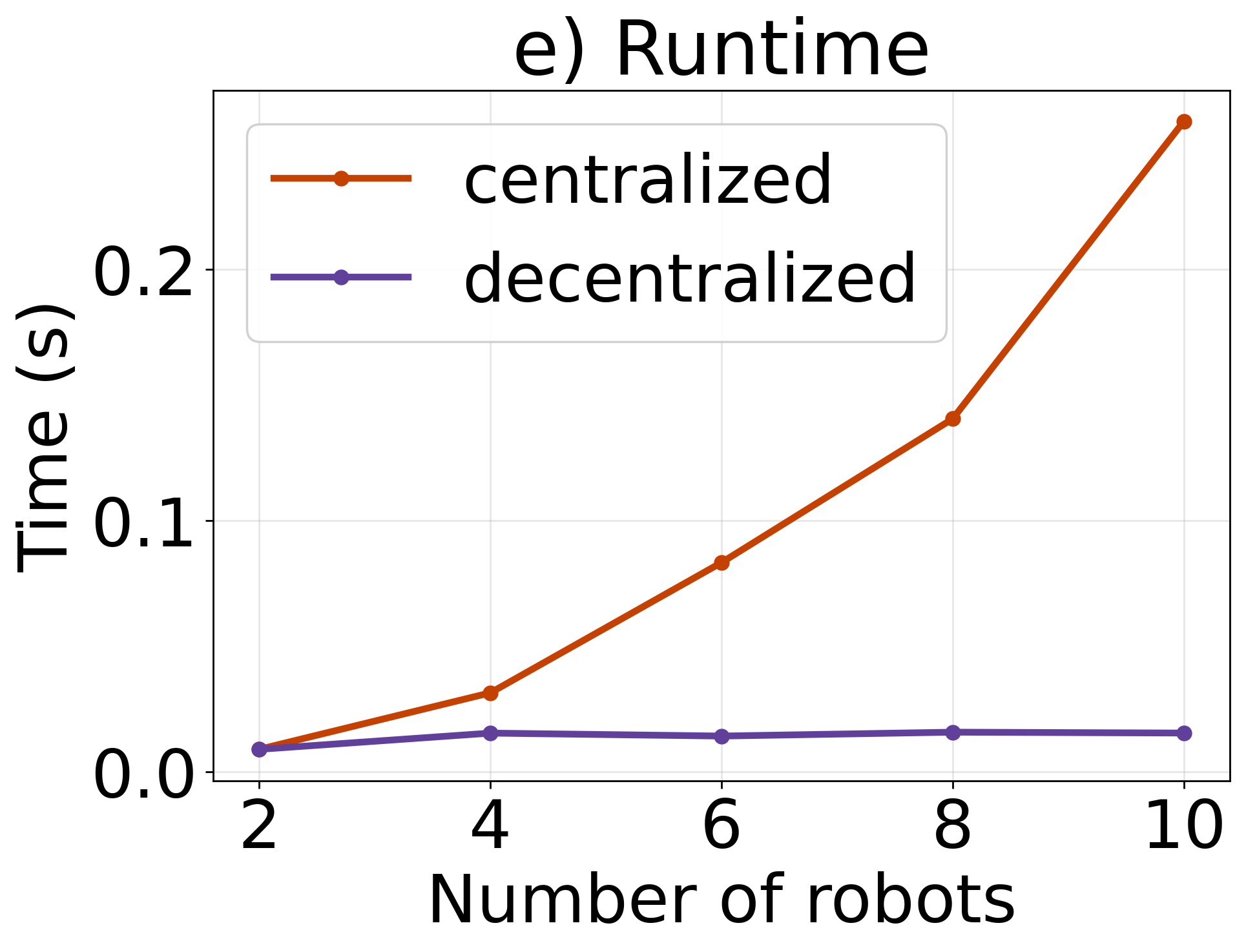}
    \end{minipage}

    \caption{Simulation performance for different team sizes.
    (a) Active fire area \(A_f(t)\), where lower values indicate less active fire.
    (b) Team-average water level \(\bar w\).
    (c) Minimum battery level across the team.
    (d) Minimum spatial safety value \(h_s(\rho)\), where higher values indicate safer operation.
    (e) Average controller runtime per iteration.
    Subplots (a)--(d) share the same legend.}
    \label{fig: Sim evaluations}
\end{figure*}

The performance evaluations for teams of \(2,4,6,8\), and \(10\) robots are shown in Fig.~\ref{fig: Sim evaluations}. The fire-area plot shows that all configurations reduce the burned area compared to the no-drone baseline. Typically, centralized control is expected to outperform decentralized control due to the available real-time global information. However, in this application, the decentralized controller \eqref{eq: decentralized model} outperforms its counterpart \eqref{eq: centralized controller} in nearly all cases, at the cost of greater resource consumption. For analysis, we interpret the suppression process as three phases: the initial response phase, the main suppression phase, and the residual cleanup phase. 

During the initial response phase, robots move towards the fire region and start deploying water. In this phase, the decentralized controller tends to reduce the peak fire area more effectively than the centralized controller. This occurs for two reasons. First, in the centralized formulation, UAVs far from the fire may stay stationary because the other robots closer to the fire can satisfy the global CLF with lower control cost. In contrast, with the decentralized controller, an isolated UAV cannot rely on teammates to satisfy the local CLF, so it is driven more aggressively towards the fire. 
Second, the centralized controller optimizes the flow rate \(f\) on the team level, preserving water for drones further away. However, the decentralized controller assumes the worst case of zero water deployment from its neighbors, causing it to deploy more water at an earlier stage.

During the main suppression phase, after most robots surround the dominant burning zone, the centralized controller eliminates the fire area faster than the decentralized one. This is reflected in Fig.~\ref{fig: Sim evaluations}(a) with a steeper slope after the fire reaches its peak. 
This is due to the efficient coordination of the entire team, which prioritizes the larger fires. However, this same prioritization delays the response to smaller hot spots, as discussed next. 

During the residual cleanup phase, the initial fire region is mostly suppressed, but smaller hot spots may have spread from it. The decentralized controller reacts to these smaller hot spots much earlier because each robot optimizes locally within its detection radius. In contrast, the centralized control prioritizes the dominant fire region as its elimination is the most rewarding. As the number of robots increases, the decentralized controller has more agents to track and suppress these smaller hot spots as they spread from the dominant region. This is why its fire extinction time improves significantly with respect to team size. The centralized controller also benefits from additional agents, but at a less noticeable magnitude due to the team-level target still prioritizing full elimination of the dominant fire region before smaller hot spots. 

The improved suppression performance of the decentralized controller comes at the cost of greater resource consumption. The lower water reservoir in Fig.~\ref{fig: Sim evaluations}(b) shows that the decentralized controller deploys more water because each robot acts on its local objective without relying on neighbor contributions.
Similarly, the minimum battery level, in Fig.~\ref{fig: Sim evaluations}(c), is always lower for the decentralized controller, as the robots begin moving earlier towards the initial fire and spreading hot spots. The centralized controller spends energy more selectively, delaying motion until they are rewarding enough on a team level. Importantly, the minimum \(E_i\) plot shows both controllers maintain sufficient energy over multiple charge cycles across all simulations. 
Figure~\ref{fig: Sim evaluations}(d) shows that the centralized controller operates closer to the boundary of \(\mathcal{A}\), where the decentralized controller acts more conservatively. Nevertheless, both controllers maintain the safety of the entire team over all simulation runs despite localization and motion uncertainty. 

Finally, subplot~(e) of Fig.~\ref{fig: Sim evaluations} demonstrates the computational advantage of the decentralized controller. 
Notably, with the parallel core implementation, where each robot is assigned to a separate CPU, the computation time stays nearly constant for all numbers of robots. In contrast, the centralized controller's computational demand increases significantly with the number of robots involved, reaching \(0.26\)~s per iteration with a team of \(10\) robots. Since a typical drone control loop requires approximately \(0.1\)~s per step, or \(10\)~Hz, the current centralized implementation becomes unsuitable for large teams, while the decentralized controller remains practical for deployment. 
Overall, the results demonstrate the effectiveness of both proposed controllers in the wildfire-suppression application under localization and motion uncertainty. The decentralized controller provides faster response and greater computational scalability at the cost of increased water and battery consumption; stronger coordination during the main suppression phase may further improve its performance.

\section{Experiment Results}
\begin{figure}[t]
    \centering
    \includegraphics[width=0.85\linewidth]{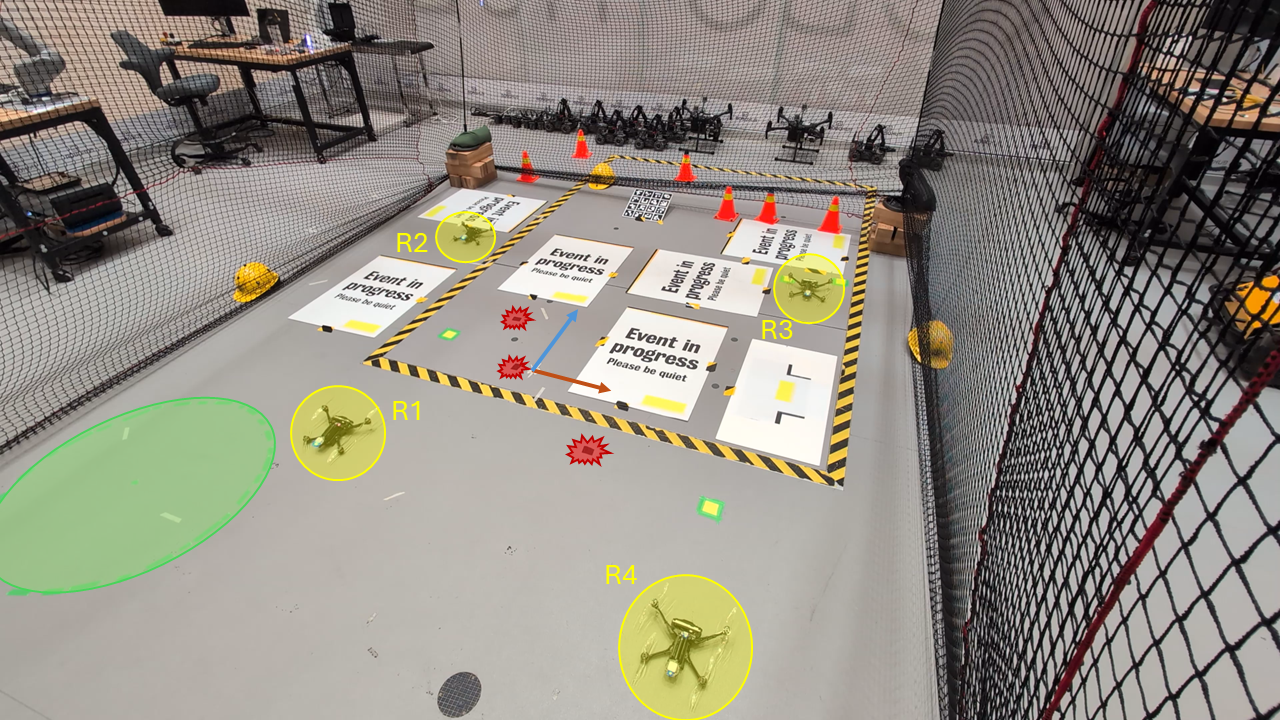}
\caption{Experimental setup with quadcopters in yellow, charging zone \(\mathcal{C}\) in green, initial fire spots in red, and red \(x\)- and blue \(y\)-axis arrows.}
    \label{fig: exp_setup}
\end{figure}
\begin{sidewaysfigure*}[p]
    \centering
    
    \setlength{\tabcolsep}{1pt}
    \renewcommand{\arraystretch}{0.9}

    \begin{adjustbox}{max width=0.8\linewidth, max height=0.8\textheight, center}
    \begin{tabular}{c c c c}

        \includegraphics[width=0.23\linewidth]{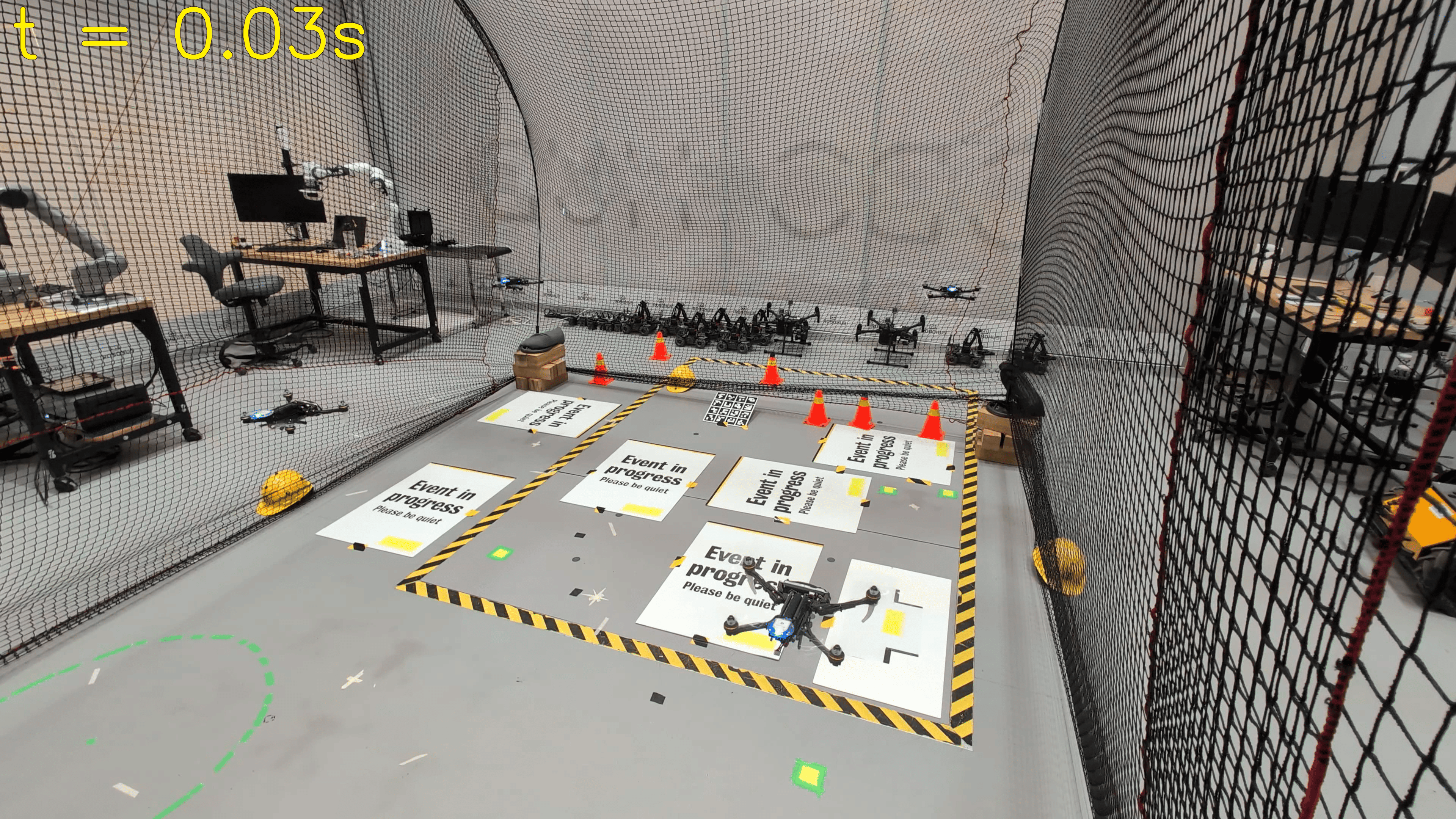} &
        \includegraphics[width=0.23\linewidth]{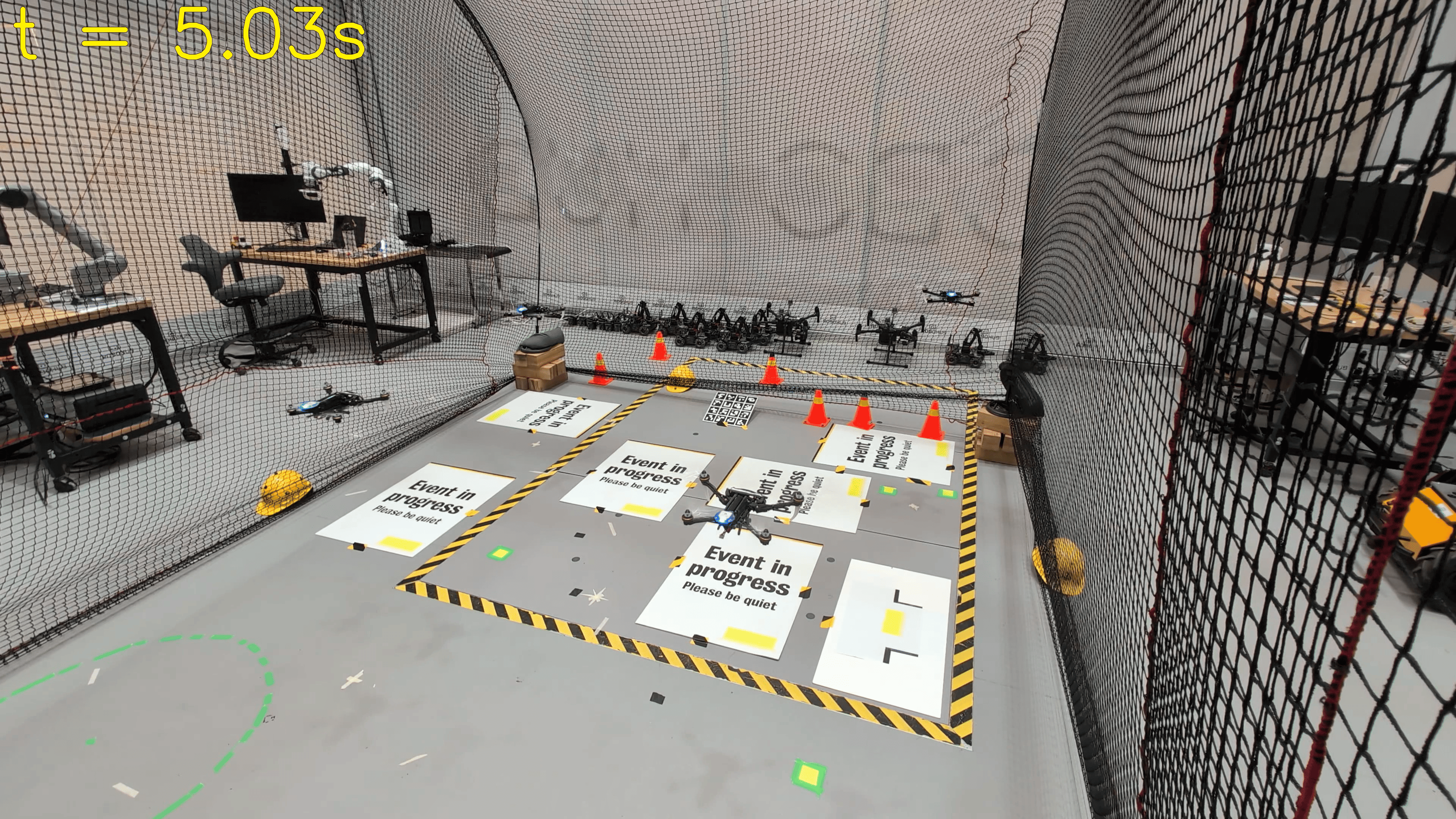} &
        \includegraphics[width=0.23\linewidth]{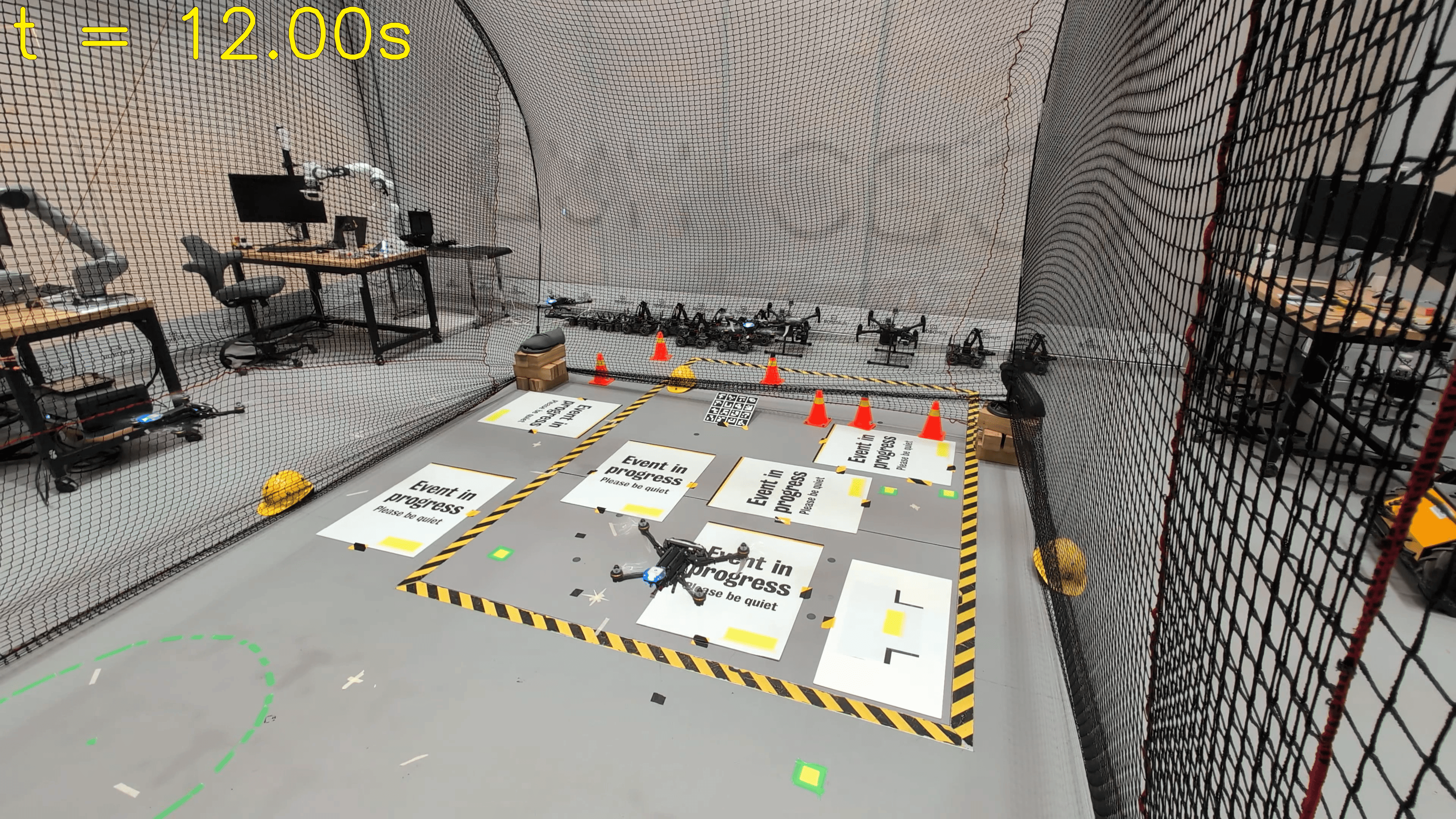} &
        \includegraphics[width=0.23\linewidth]{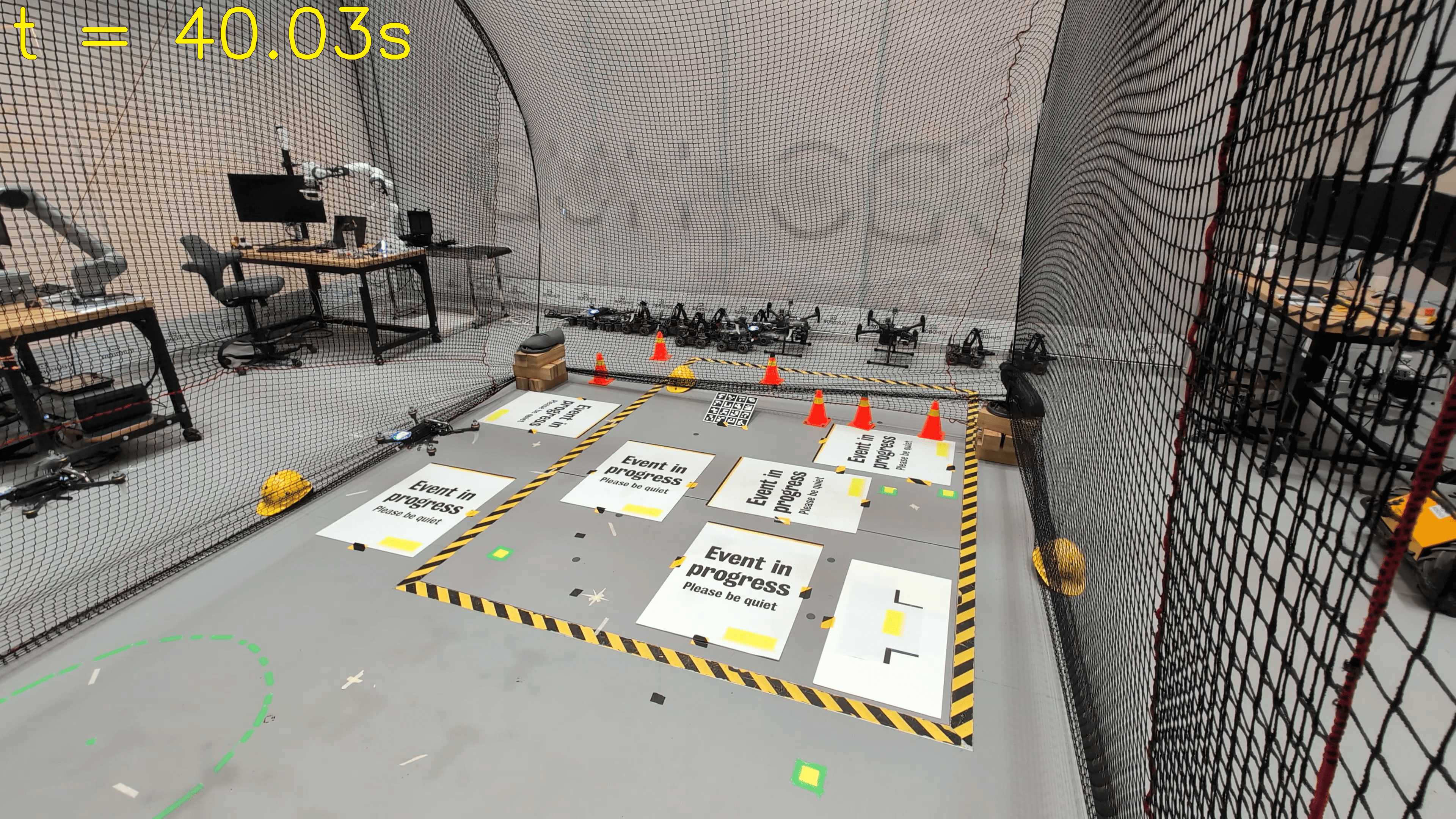} \\[-0.2em]

        \includegraphics[width=0.23\linewidth]{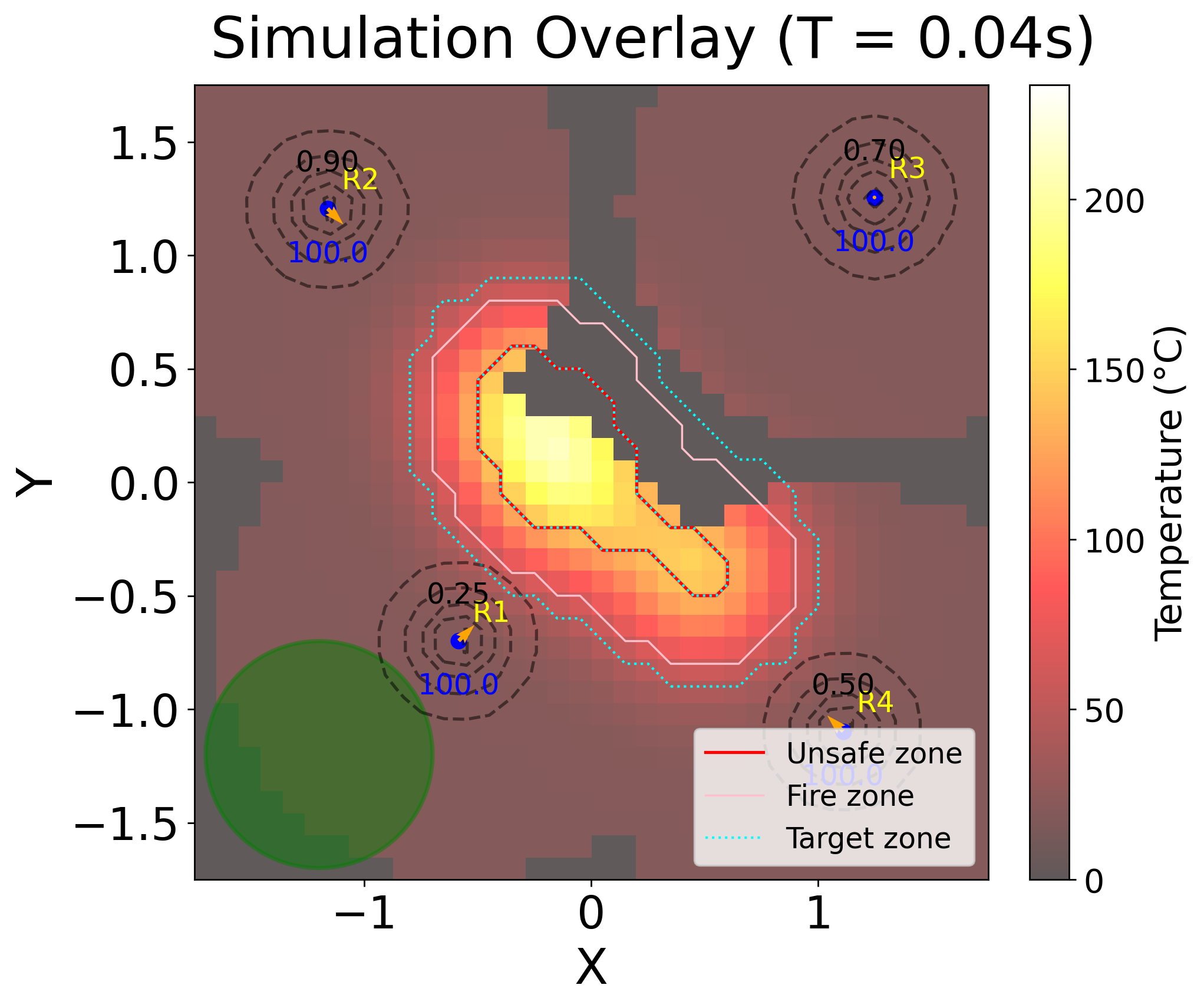} &
        \includegraphics[width=0.23\linewidth]{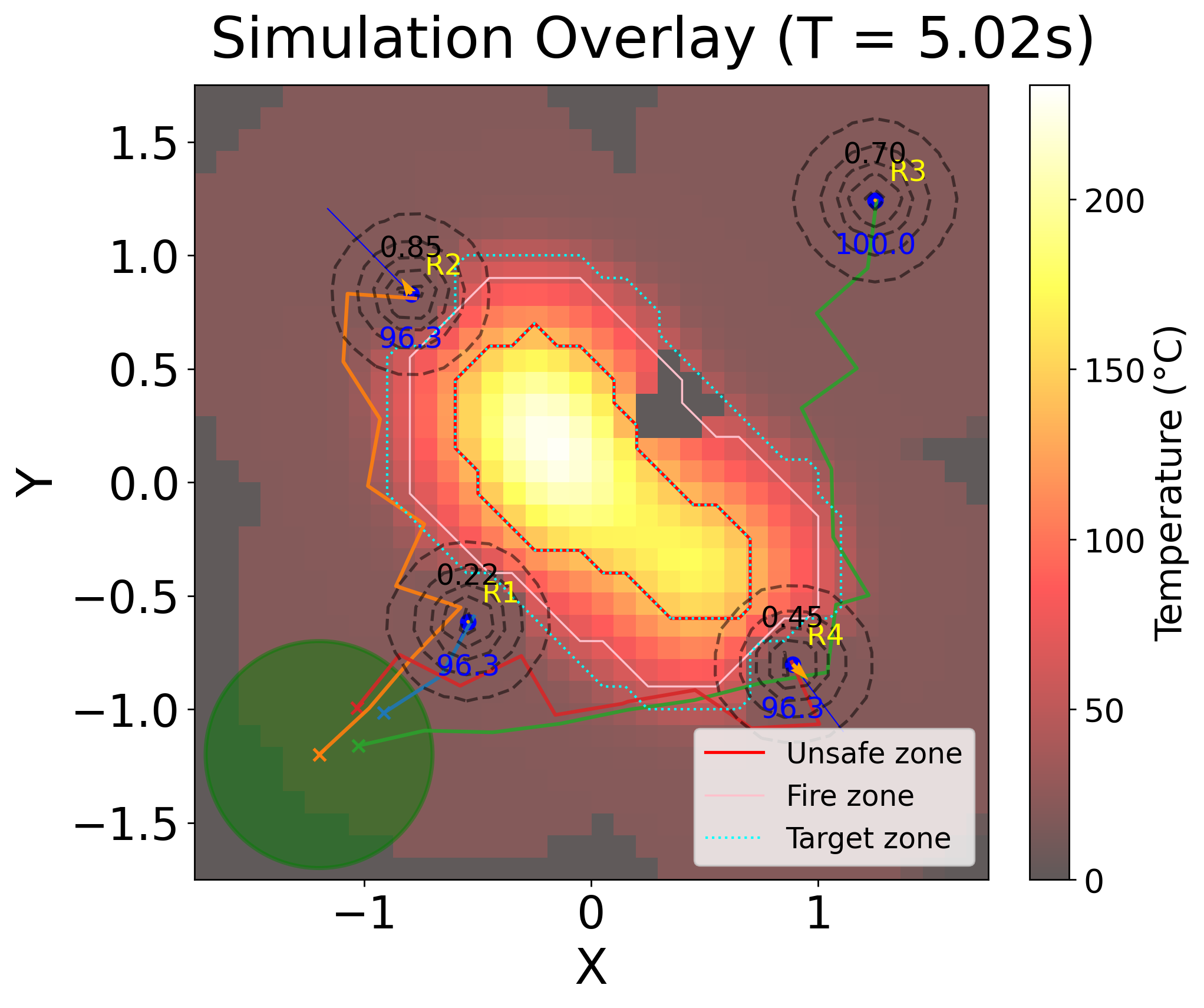} &
        \includegraphics[width=0.23\linewidth]{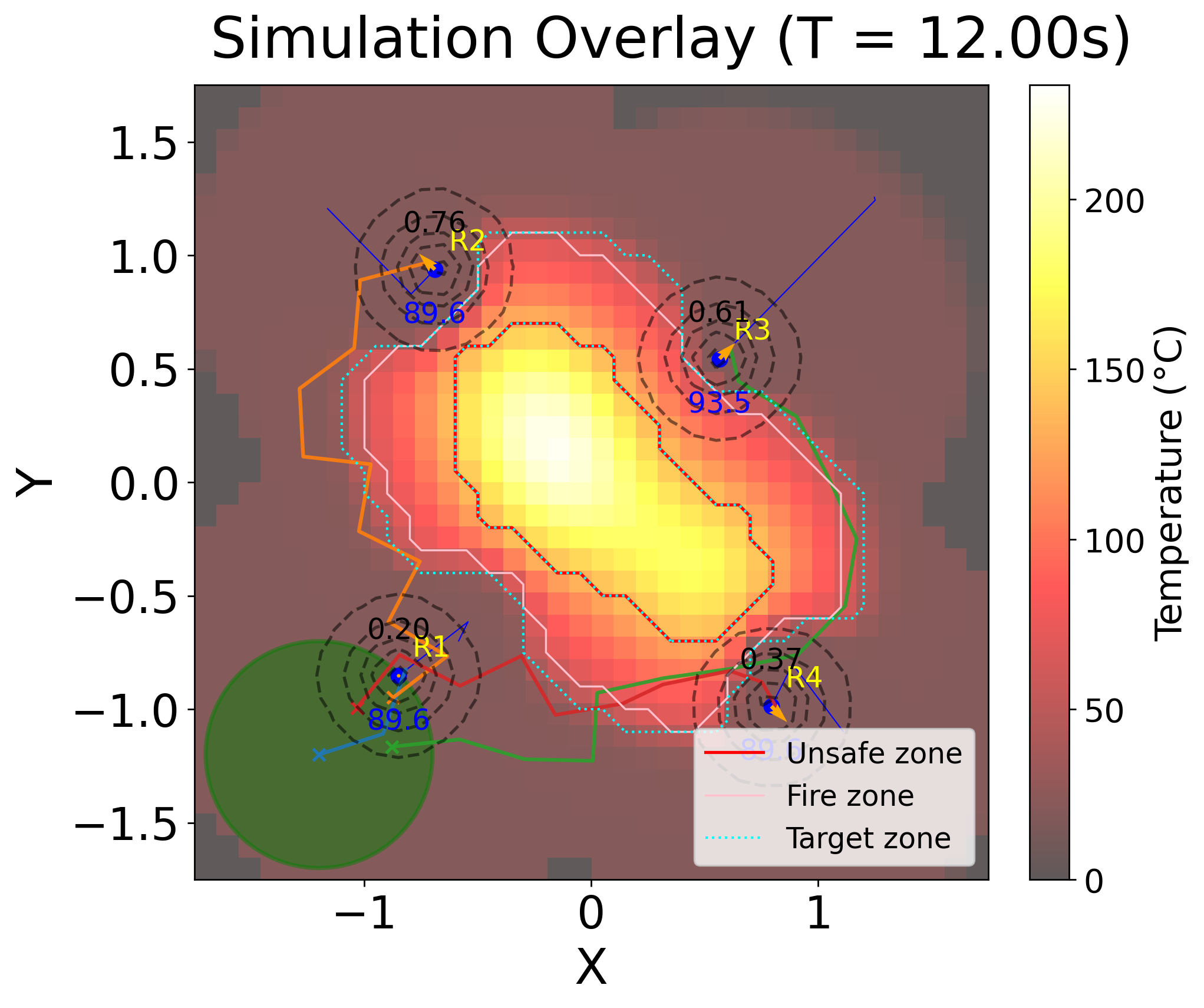} &
        \includegraphics[width=0.23\linewidth]{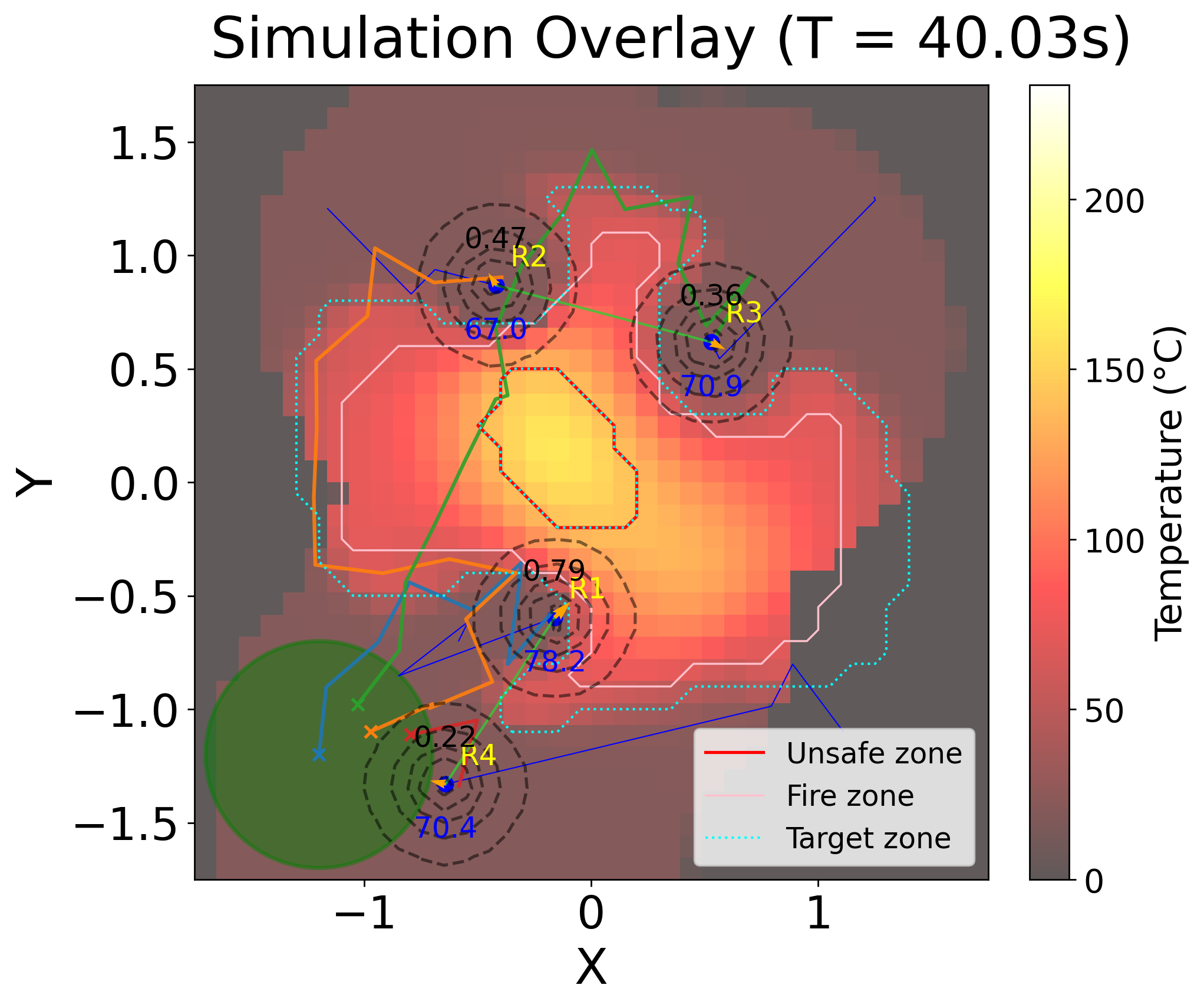} \\[0.8em]

        \includegraphics[width=0.23\linewidth]{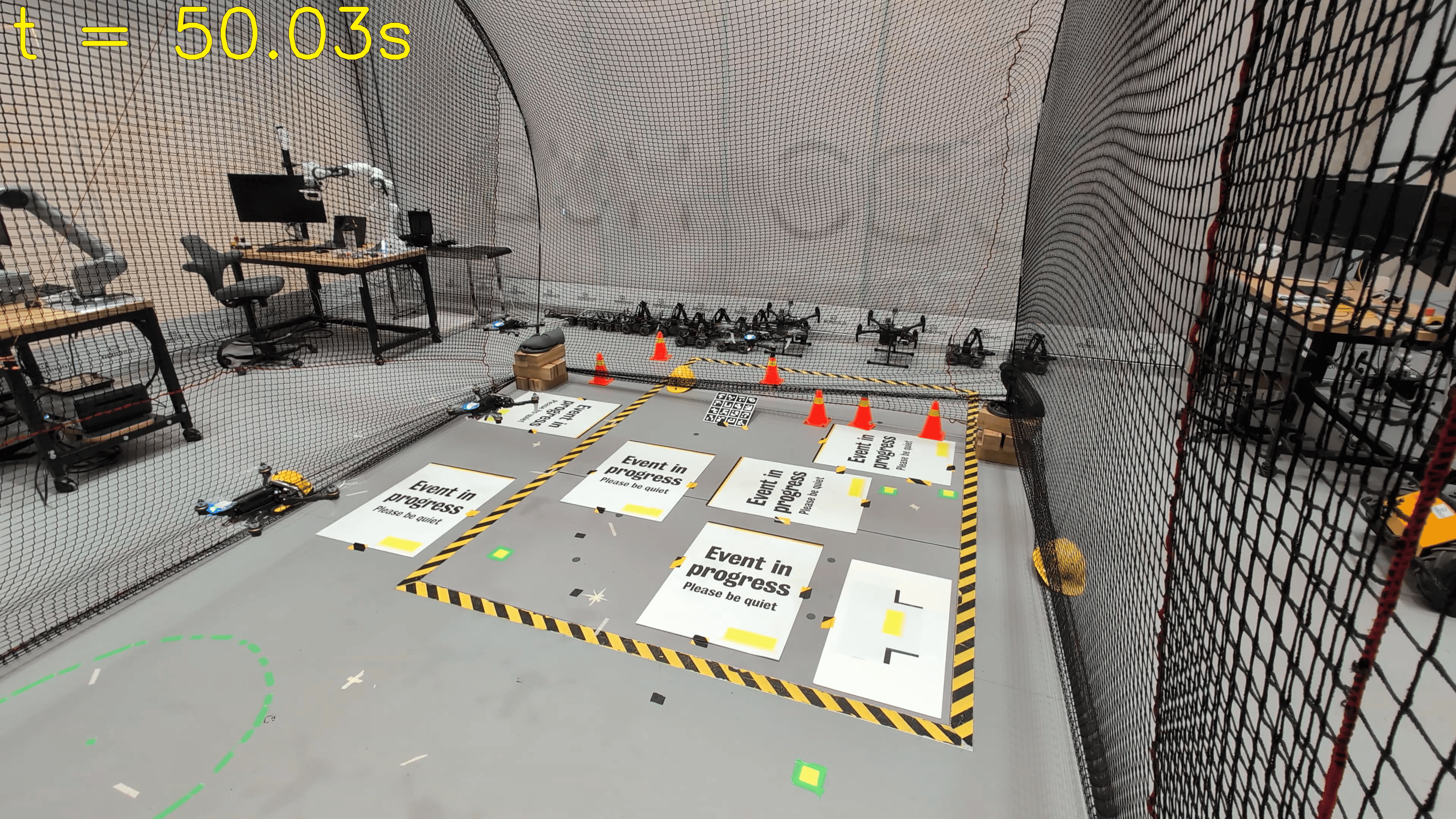} &
        \includegraphics[width=0.23\linewidth]{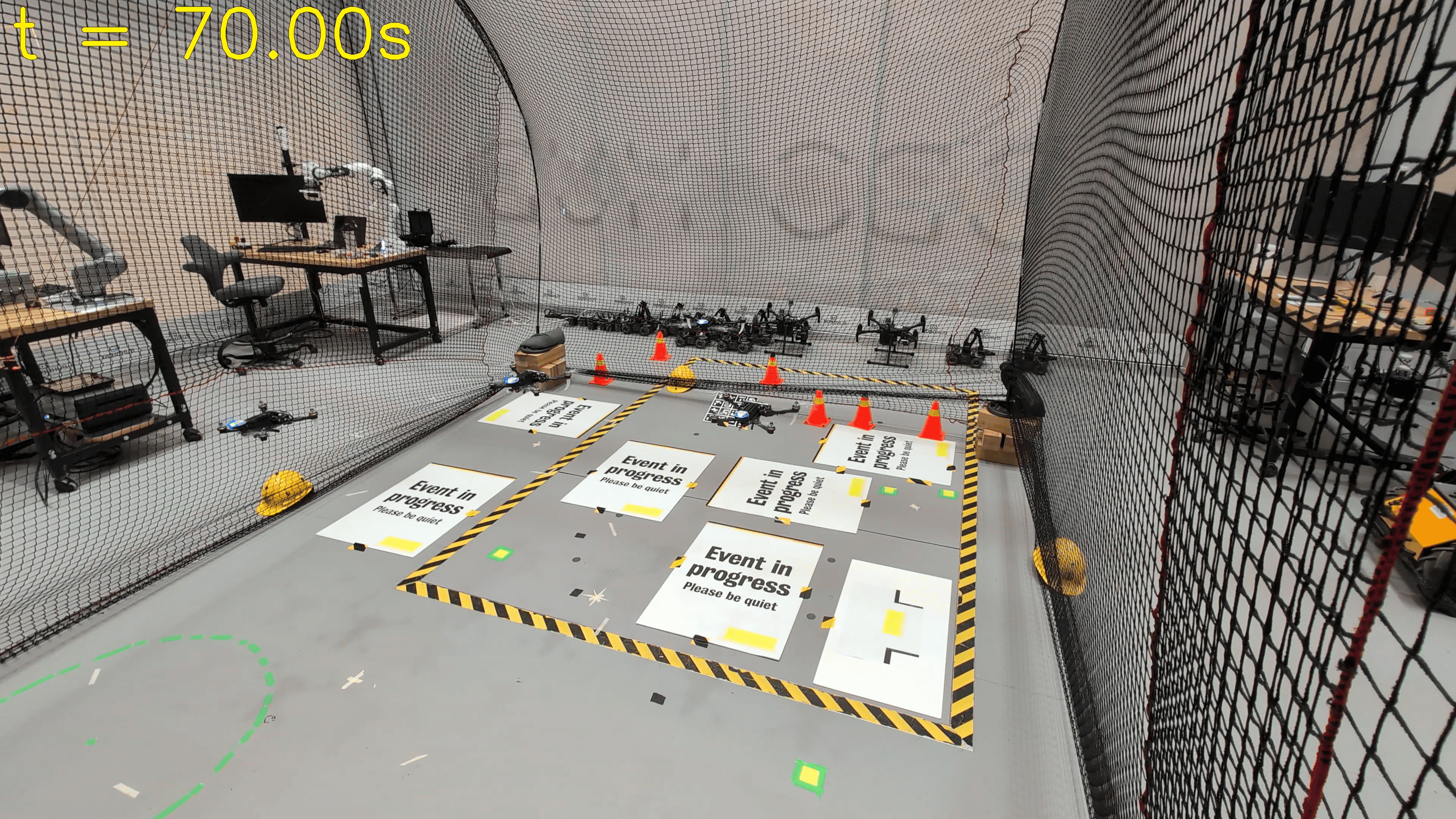} &
        \includegraphics[width=0.23\linewidth]{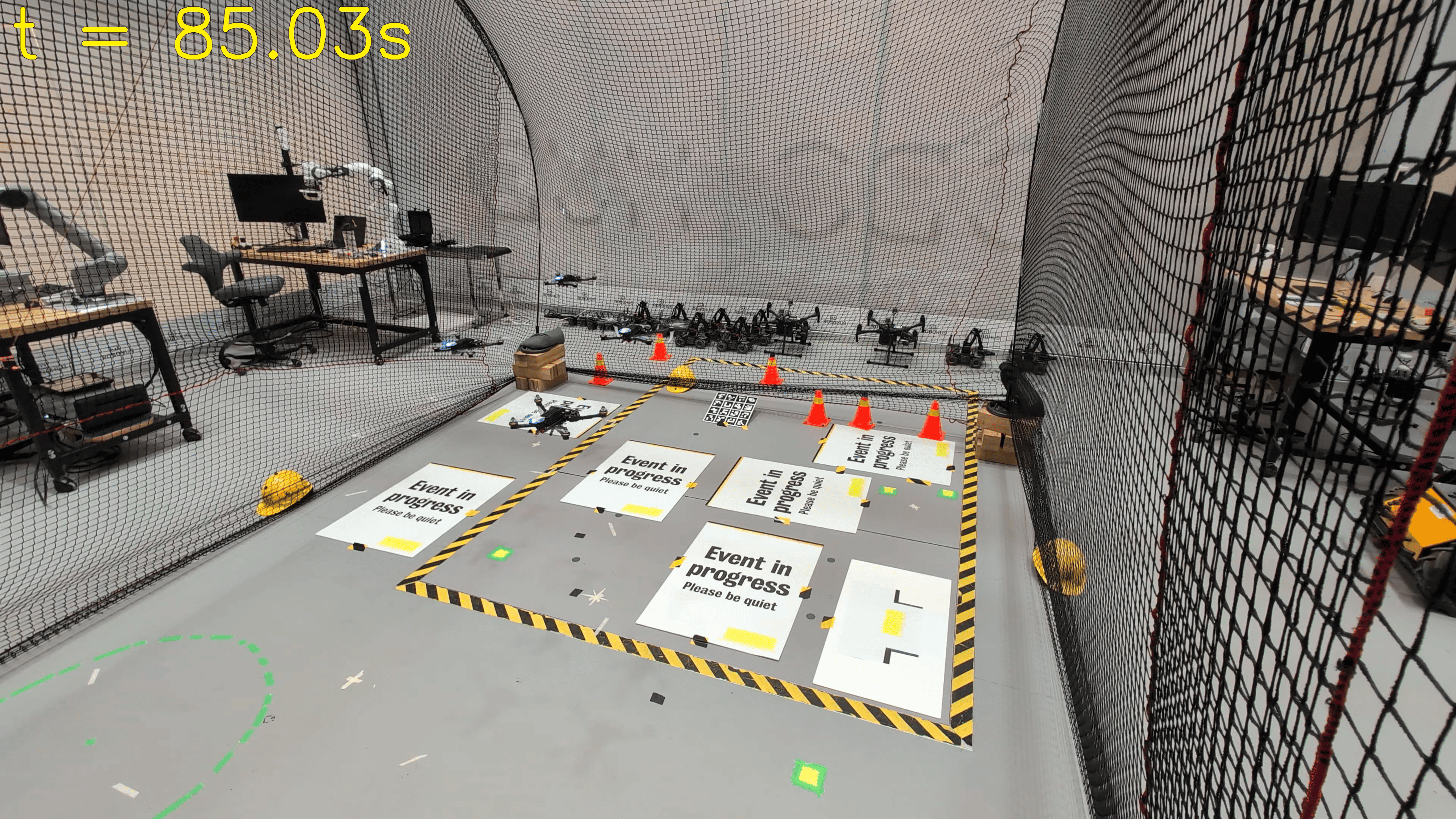} &
        \includegraphics[width=0.23\linewidth]{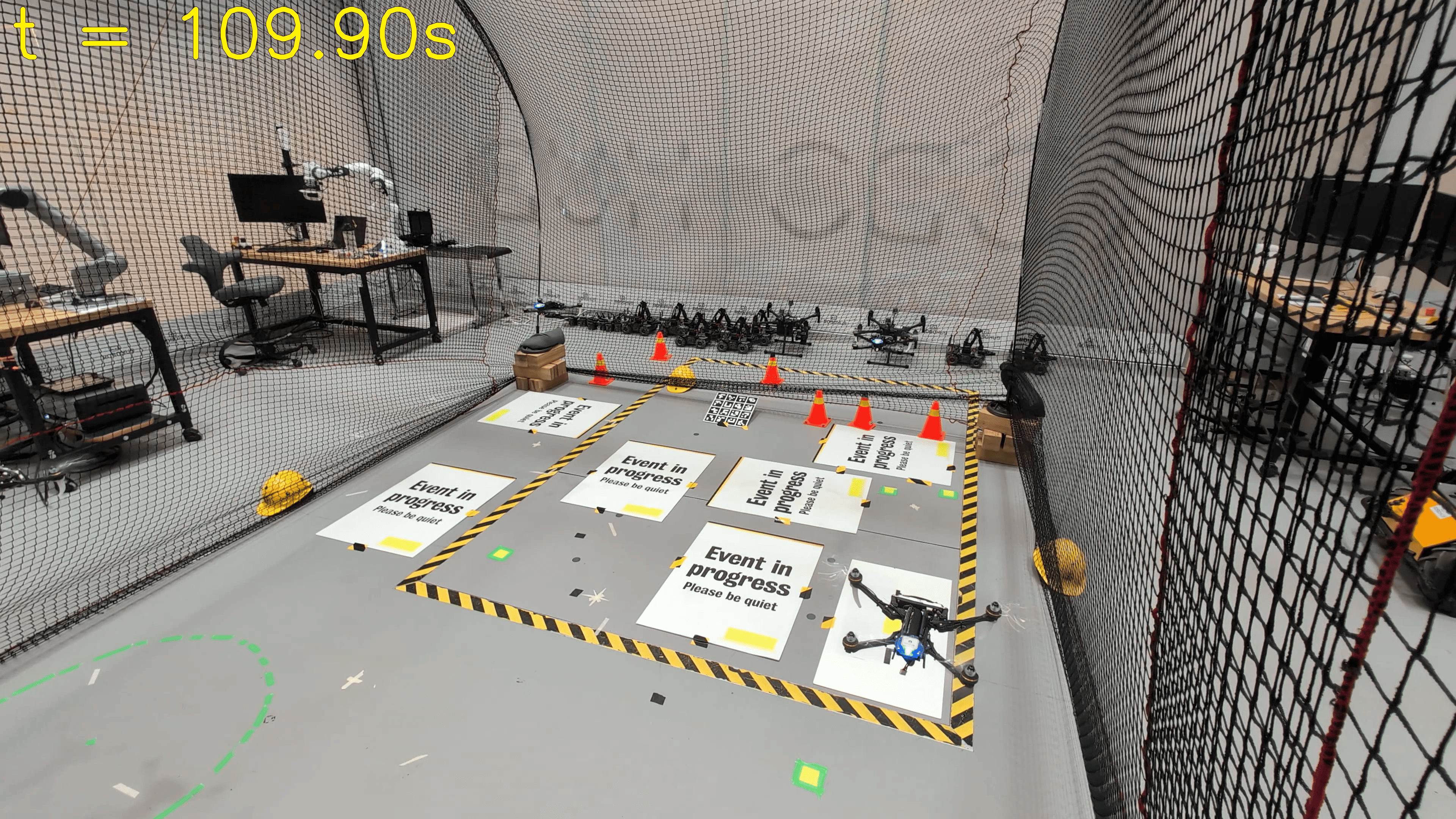}\\[-0.2em]

        \includegraphics[width=0.23\linewidth]{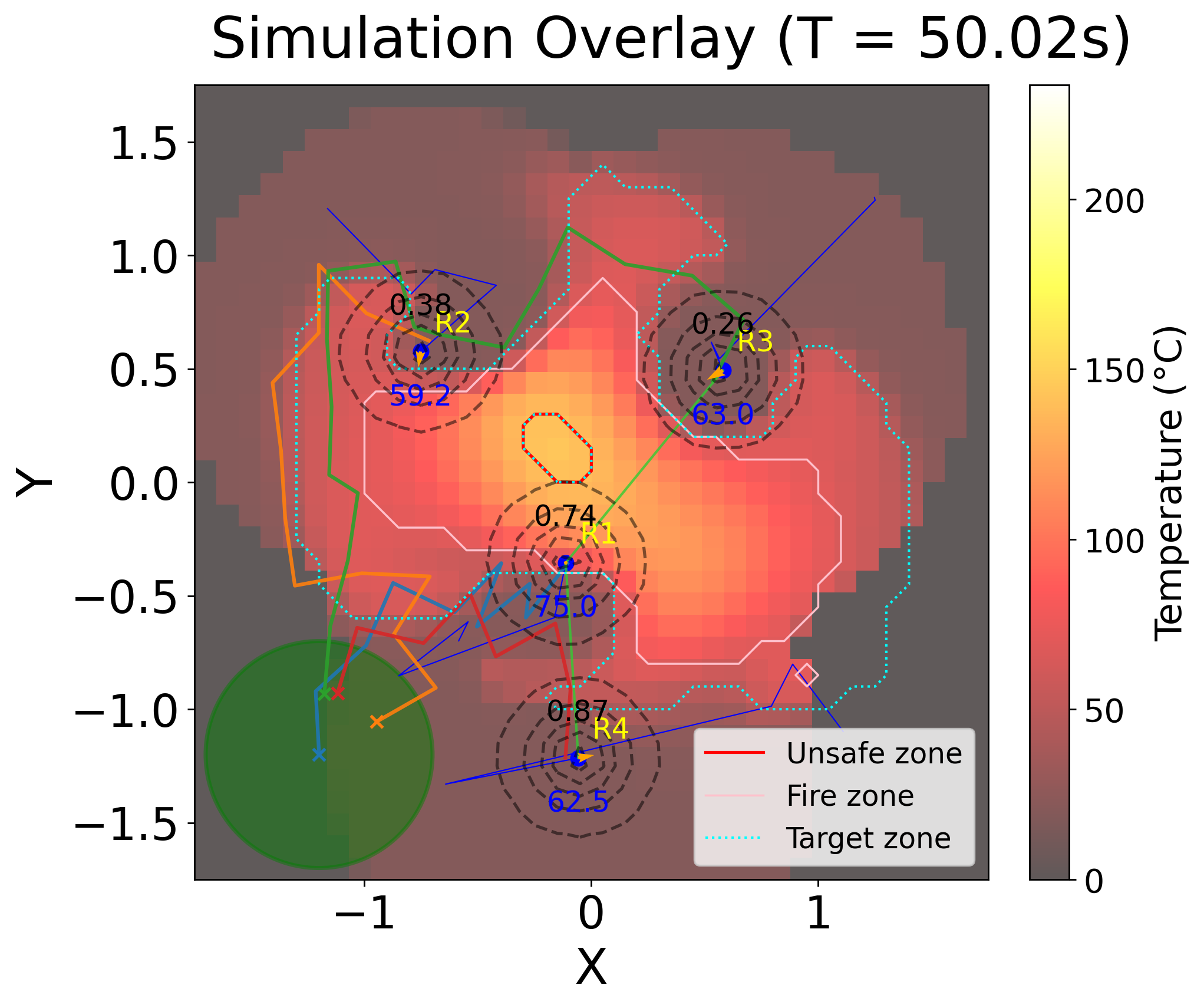} &
        \includegraphics[width=0.23\linewidth]{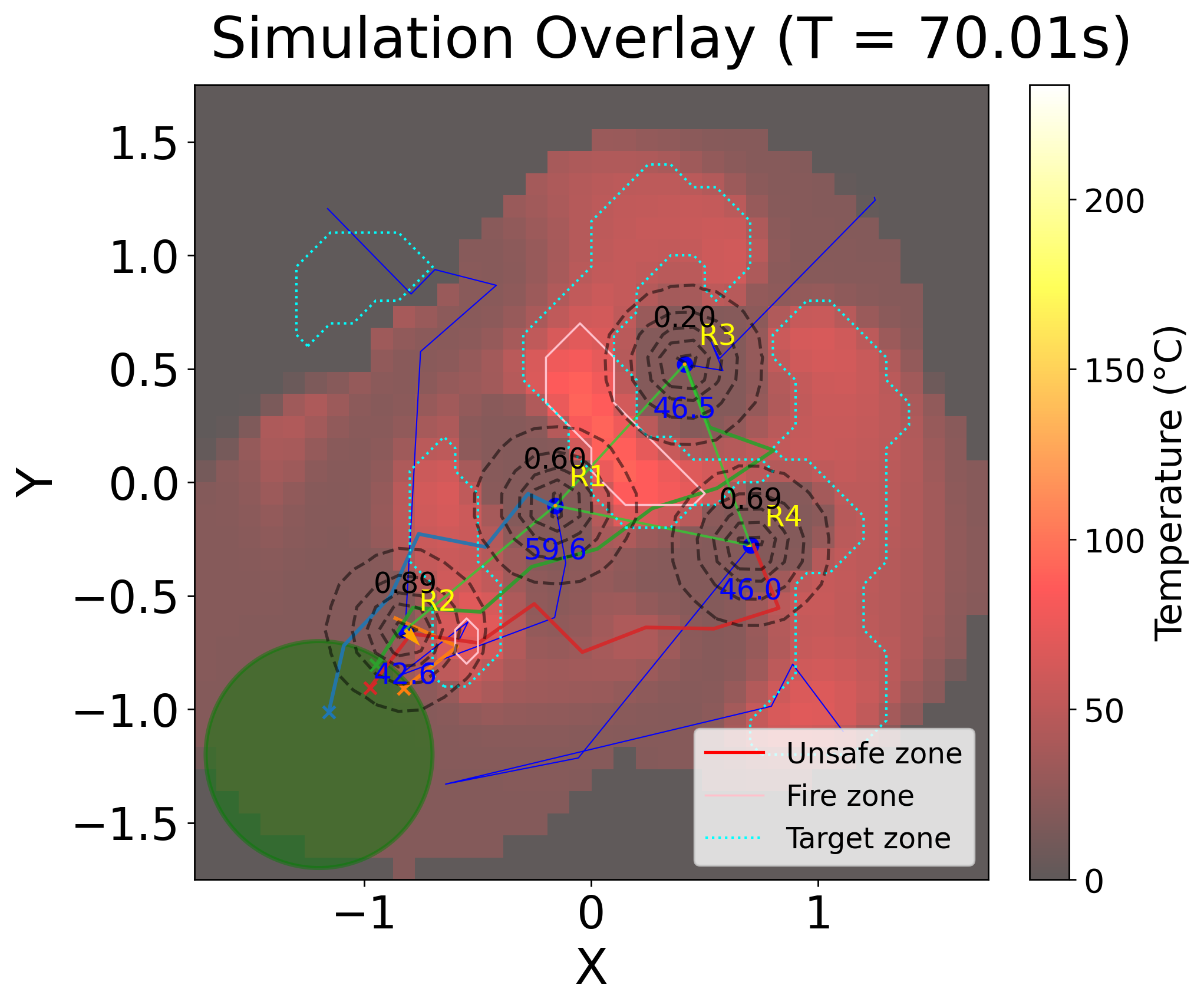} &
        \includegraphics[width=0.23\linewidth]{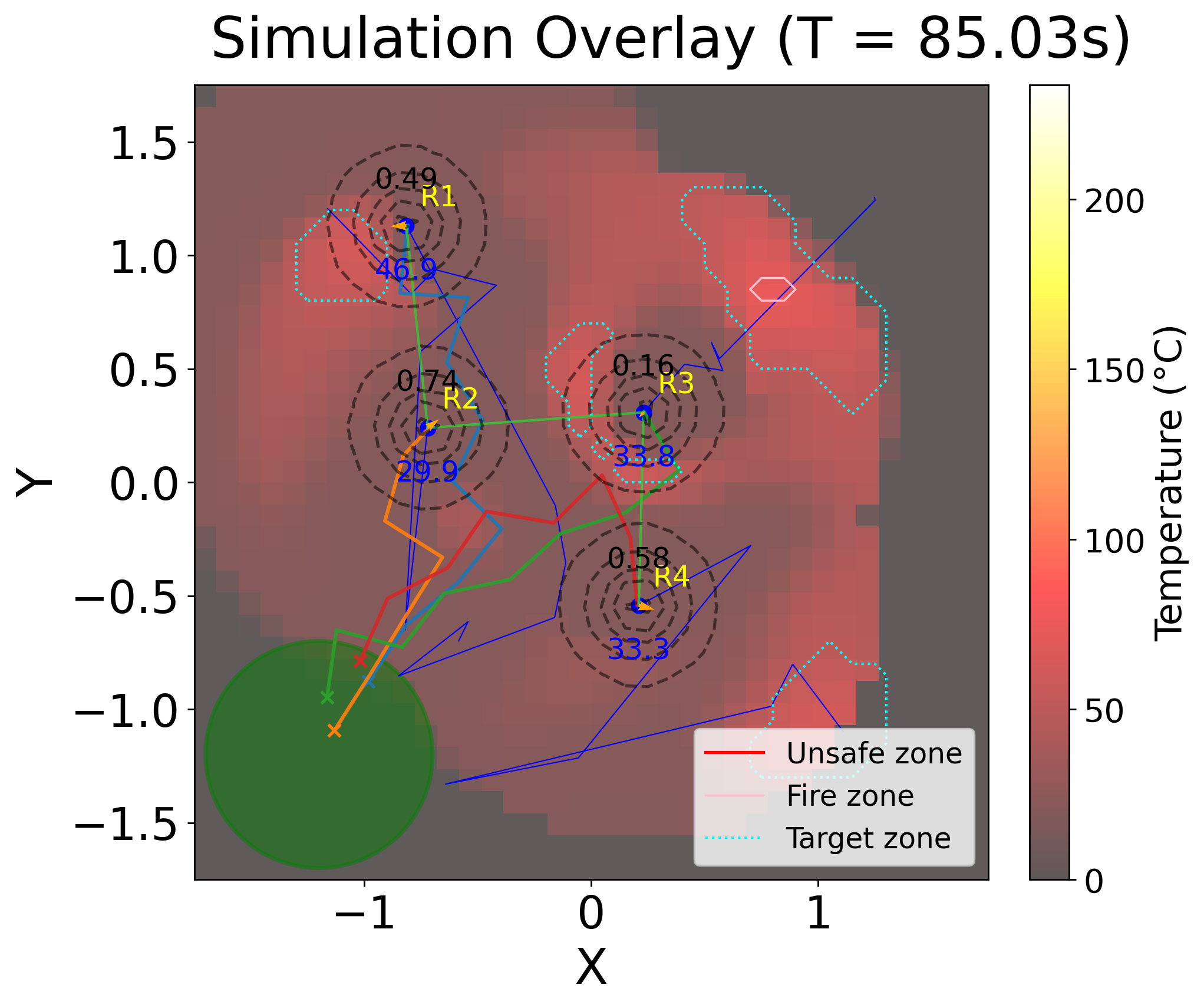} &
        \includegraphics[width=0.23\linewidth]{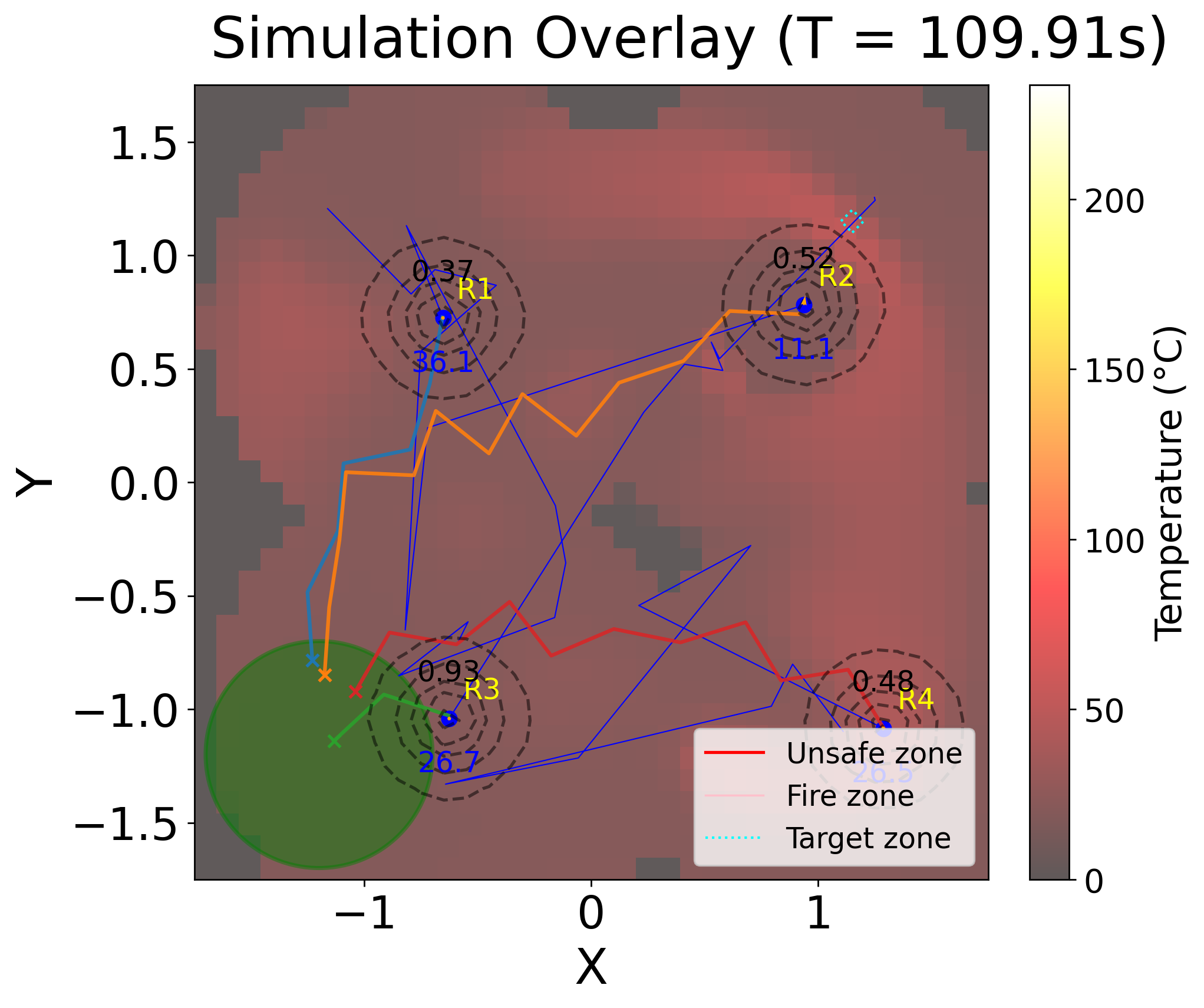}
    \end{tabular}
    \end{adjustbox}

    \caption{Time sequence of the experiment and corresponding simulation representation. 
    For each time group, the top row shows the experiment with four quadcopters, and the bottom row shows the simulation overlay. 
    Robot positions \(x_i\), robot PDFs \(\rho_i\), control inputs \(u_i\), previous trajectories, and planned safe paths to the green charging zone \(\mathcal{C}\) are shown in blue, black contours, gold, blue, and distinct path colors, respectively. 
    The battery level \(E_i\), water level \(w_i\), and robot ID \(i\) are displayed above each robot in black, blue, and gold, respectively. 
    The detected temperature field is shown as the background heatmap, with the unsafe zone \(\mathcal{A}\) in red, active-fire zone \(\mathcal{F}\) in pink, and deployment zone \(\mathcal{D}\) in cyan.}
    \label{fig:exp_robot_timeSequence}
\end{sidewaysfigure*}

To evaluate the performance of the decentralized controller in practice, experiments with four Modal AI Seeker quadcopters were conducted in a $3.5 \times 3.5$~m cage with the same fire and charger setup as the simulations.
Each quadcopter has a detection radius of $1.1$~m and a $0.7$~m collision radius, as a safety buffer beyond the physical tip-to-tip distance of $0.5$~m. The setup is shown in Fig.~\ref{fig: exp_setup}. The floor markers provide localization features for the onboard cameras, while the quadcopter downwash introduces noticeable motion and localization noise. Each quadcopter computes its own control input using only local sensing and neighbor map sharing, mimicking deployment.
The time sequence of the experiment is shown in Fig.~\ref{fig:exp_robot_timeSequence}, with the supplementary video attached and available online at \href{https://youtu.be/LrXBq0-Ez10}{https://youtu.be/LrXBq0-Ez10}.

Initially, R3 is isolated and far from the fire, so it remains stationary as its locally detected temperature field contains no fire. By \(t = 5.02\)~s, R2 and R4 move toward the fire to deploy water, while retreating from \(\mathcal{A}\) to avoid possible thermal damage. R1 remains near the charger \(\mathcal{C}\) because of its low initial battery level. 
Once the fire spreads into R3's detection radius, R3 moves towards the fire for suppression by \(t = 12\)~s. At the same time, R1 reaches the charger, while the other two drones move laterally along the target zone \(\mathcal{D}\).  Around \(t = 40\)~s, R4 moves towards \(\mathcal{C}\) to ensure battery sufficiency. Through the map shared by R4, R1 moves towards the right side of the map. R2 moves towards R3 to suppress the fire between them. 
At \(t = 50\)~s, R2 leaves the neighborhood of R3 as it follows the orange safe path to the charger while continuing to deploy water. R1 then becomes R3's neighbor, and the two drones suppress the main burning zone together. At the same time, R4 moves to the right of the map, where the local target PDF is higher due to a lack of information on the left side. By \(t = 70\)~s, R2 is recharged and suppresses a small hot spot within its detection radius, while the other 3 robots work together to eliminate the central fire zone. 
By \(t = 85\)~s, the initial fire zone has been eliminated. R3 attempts to move towards the charger; however, this motion is blocked due to collision avoidance with R1 and R4. R1 tracks the remaining deployment zone on the top left, while R2 rejoins the group and receives information of \(\mathcal{D}\) on the top right from R3's map sharing. R4 moves towards the bottom right hot spot after eliminating the central fire. 
In the end, by \(t = 110\)~s, the field temperature has dropped below the deployment threshold over nearly the entire domain, leading robots to rest. 

\begin{figure}[t]
    \centering
    \includegraphics[width=0.32\linewidth]{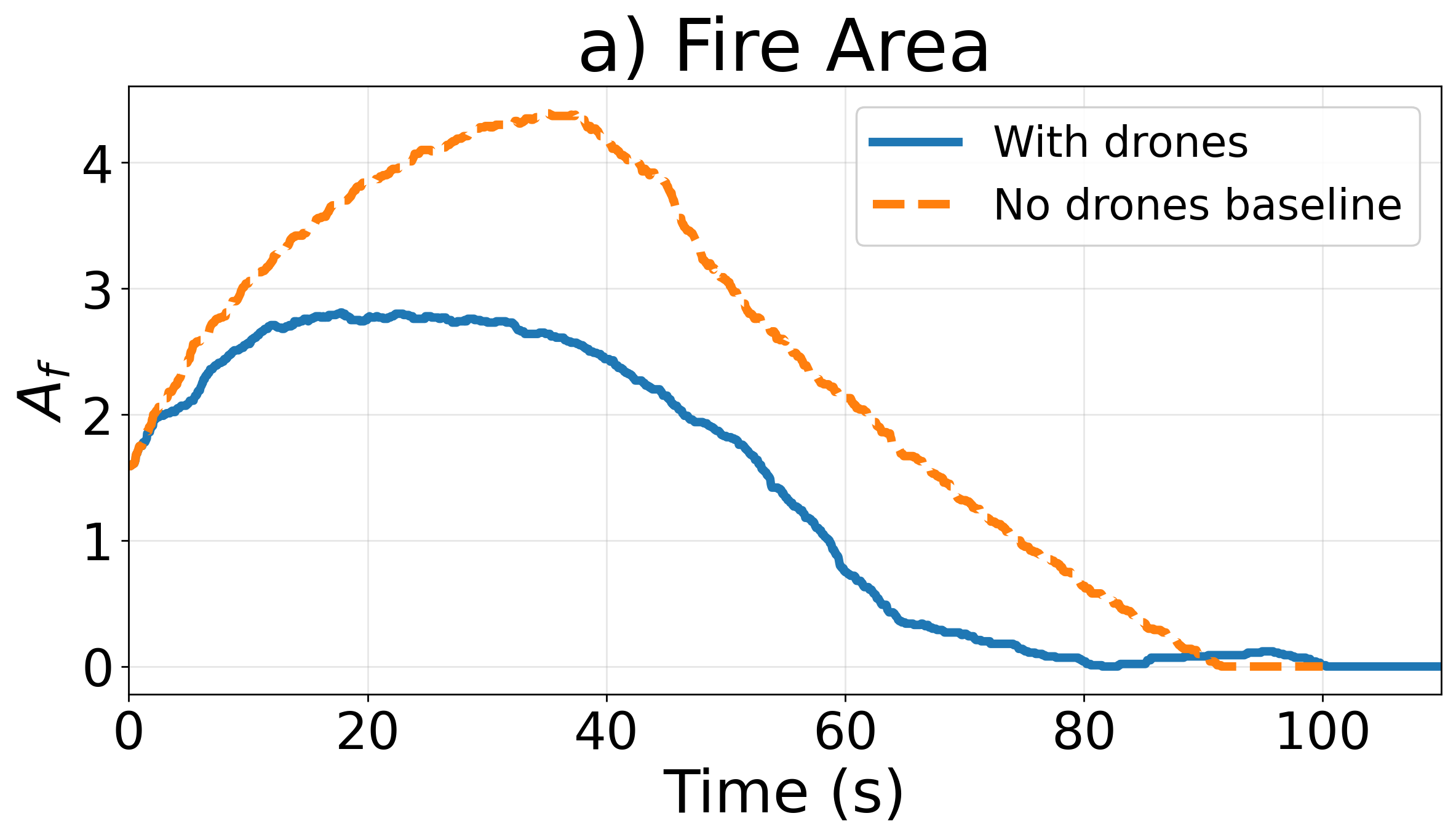}
    \includegraphics[width=0.32\linewidth]{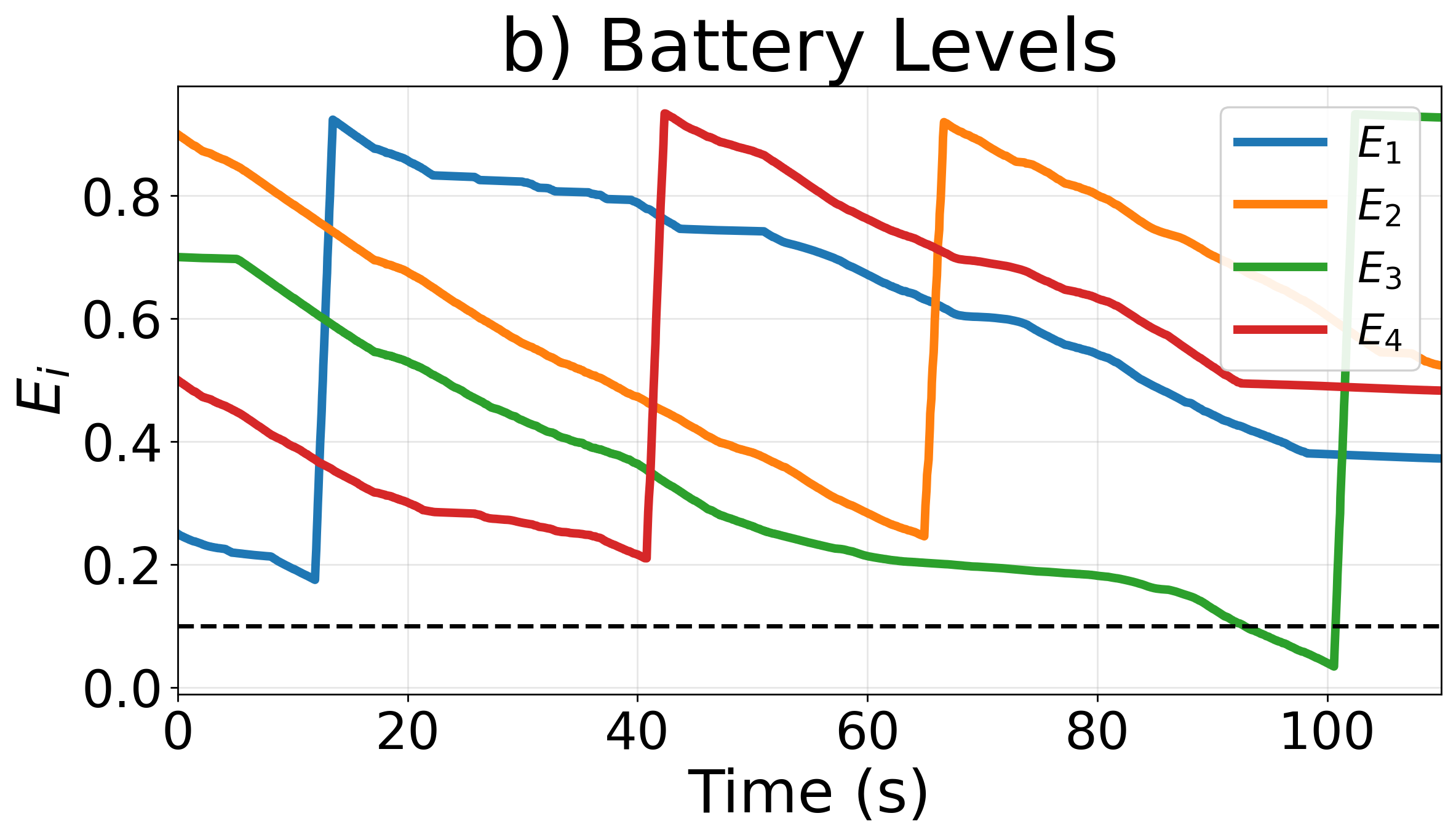}
    \includegraphics[width=0.32\linewidth]{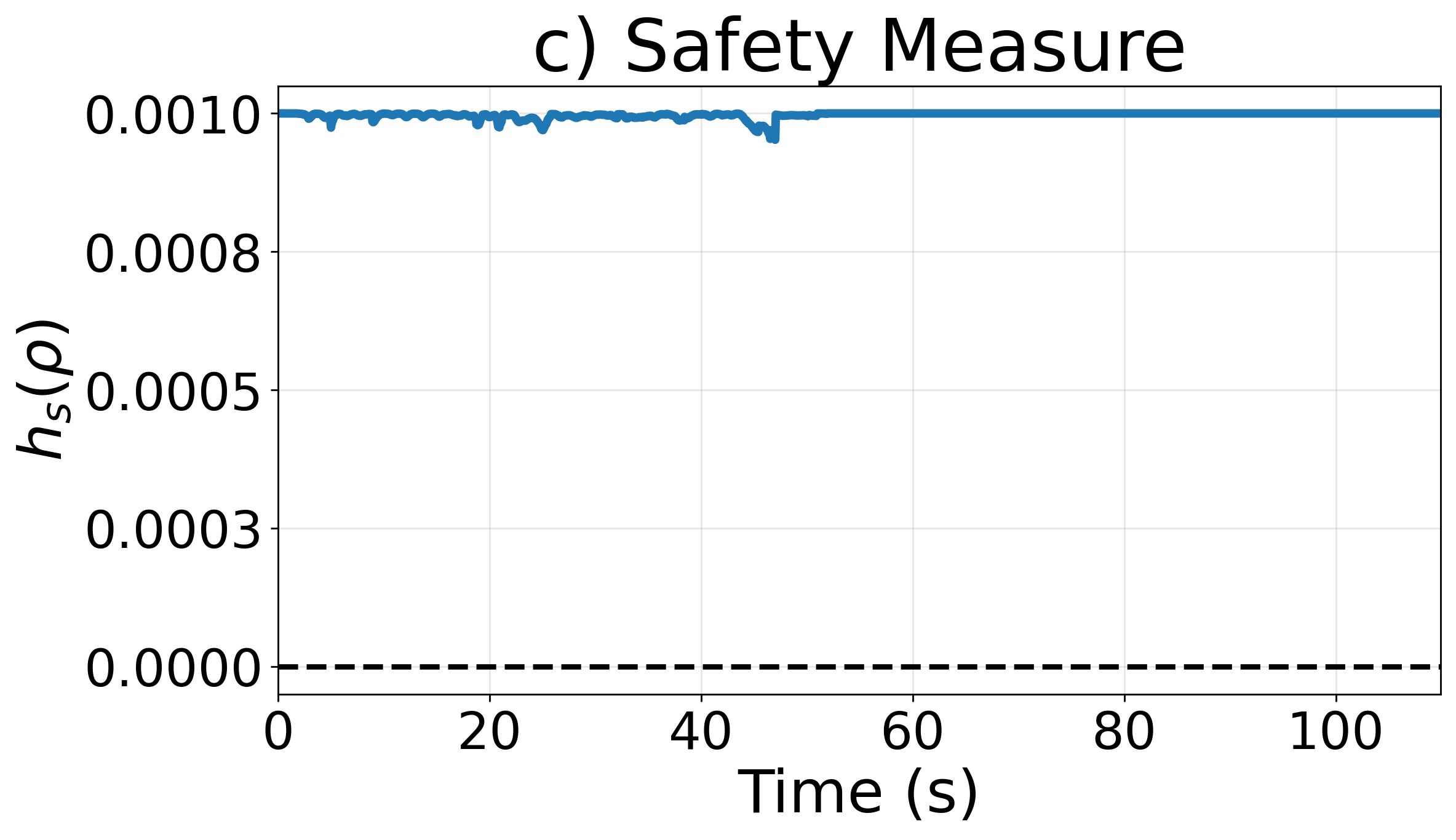}
    
    \caption{Experiment performance plots.
    (a) Active fire area \(A_f(t)\), where lower values indicate less active fire.
    (b) Individual battery level \(E_i\).
    (c) Spatial safety value \(h_s(\rho)\), where higher values indicate safer operation.}
    \label{fig: exp evaluations}
\end{figure}
The performance metrics are shown in Fig.~\ref{fig: exp evaluations}. The decentralized deployment nearly halves the peak fire area, demonstrating the effectiveness of fire suppression using only local sensing and neighbor information. Energy sufficiency is also maintained throughout the experiment, with the only violation arising from collision avoidance. In practical outdoor deployments, this issue can be eliminated by assigning drones to different flight altitudes. The spatial safety metric remains high throughout the experiment, confirming that the team remains outside of the unsafe thermal region while still achieving fire suppression. 
In conclusion, these results demonstrate the practical, real-time feasibility of the proposed decentralized controller despite localization and motion uncertainty. 


\section{Conclusion}
In this paper, we introduced a decentralized density-based PDE-constrained control framework for multi-robot wildfire suppression applications. The proposed method combines fire-aware target generation, local information sharing, water deployment, spatial safety, and energy-awareness within a unified optimization-based controller. Simulation and quadcopter experiment results show that the robots can reduce fire spread while respecting safety and energy constraints under limited sensing and communication.
Future work will incorporate water-refill planning and improve coordination during the main suppression phase. 

%
%
\bibliographystyle{splncs04}
\bibliography{references}

@article{Archer2004DynamicalDF,
  title={Dynamical density functional theory for interacting Brownian particles: stochastic or deterministic?},
  author={Andrew J. Archer and Markus Rauscher},
  journal={Journal of Physics A},
  year={2004},
  volume={37},
  pages={9325-9333},
  doi={10.1088/0305-4470/37/40/001}
}

@article{ANNUNZIATO2013487,
title = {A Fokker–Planck control framework for multidimensional stochastic processes},
journal = {Journal of Computational and Applied Mathematics},
volume = {237},
number = {1},
pages = {487-507},
year = {2013},
issn = {0377-0427},
doi = {10.1016/j.cam.2012.06.019},
author = {M. Annunziato and A. Borzì}}

@ARTICLE{RalPaper,
 author={Niu, Longchen and Notomista, Gennaro},
  journal={IEEE Robotics and Automation Letters}, 
  title={Decentralized Density Control of Multi-Robot Systems Using PDE-Constrained Optimization}, 
  year={2025},
  volume={10},
  number={4},
  pages={4045-4052},
  doi={10.1109/LRA.2025.3548501}}

@article{rrt_paper,
title = {Completeness of randomized kinodynamic planners with state-based steering},
journal = {Robotics and Autonomous Systems},
volume = {89},
pages = {85-94},
year = {2017},
issn = {0921-8890},
doi = {10.1016/j.robot.2016.12.002},
url = {https://www.sciencedirect.com/science/article/pii/S0921889015302190},
author = {Stéphane Caron and Quang-Cuong Pham and Yoshihiko Nakamura}
}

@article{ais_paper,
author = {Niu, Longchen and Notomista, Gennaro},
title = {“It Is Much Safer to Be Sparse than Connected”: Safe Control of Robotic Swarm Density Dynamics with PDE Optimization with State Constraints},
journal = {Advanced Intelligent Systems},
volume = {8},
number = {7},
pages = {e202501223},
doi = {10.1002/aisy.202501223},
url = {https://advanced.onlinelibrary.wiley.com/doi/abs/10.1002/aisy.202501223},
eprint = {https://advanced.onlinelibrary.wiley.com/doi/pdf/10.1002/aisy.202501223},
year = {2026}
}

@misc{ITSC_Energy_arxiv,
      title={Safe and Energy-Aware Multi-Robot Density Control via PDE-Constrained Optimization for Long-Duration Autonomy}, 
      author={Longchen Niu and Andrew Nasif and Gennaro Notomista},
      year={2026},
      eprint={2604.15524},
      archivePrefix={arXiv},
      primaryClass={eess.SY},
      url={https://arxiv.org/abs/2604.15524}, 
}

@inproceedings{ITSC_Energy,
  author    = {Longchen Niu and Andrew Nasif and Gennaro Notomista},
  title     = {Safe and Energy-Aware Multi-Robot Density Control via PDE-Constrained Optimization for Long-Duration Autonomy},
  booktitle = {2026 IEEE 29th International Conference on Intelligent Transportation Systems (ITSC)},
  year      = {2026},
  address   = {Naples, Italy},
  month     = sep,
  note      = {To appear}
}

@INPROCEEDINGS{MRS_paper,
  author={Niu, Longchen and Notomista, Gennaro},
  booktitle={2025 IEEE International Symposium on Multi-Robot and Multi-Agent Systems (MRS)}, 
  title={Safe Decentralized Density Control of Multi-Robot Systems using PDE-Constrained Optimization with State Constraints}, 
  year={2025},
  volume={},
  number={},
  pages={1-7},
  doi={10.1109/MRS66243.2025.11357273}}

@INPROCEEDINGS{Fire_PDE_Boundary,
  author={Belhadjoudja, M. C. and Maghenem, M. and Witrant, E. and Georges, D.},
  booktitle={2025 IEEE 64th Conference on Decision and Control (CDC)}, 
  title={Boundary Control for Wildfire Mitigation}, 
  year={2025},
  volume={},
  number={},
  pages={145-150},
  doi={10.1109/CDC57313.2025.11312804}}

@INPROCEEDINGS{Fire_PDE_OC,
  author={Georges, Didier},
  booktitle={2025 IEEE 64th Conference on Decision and Control (CDC)}, 
  title={Wildfire Mitigation Using An Aerial Firefighting Vehicle: An Optimal Control Approach}, 
  year={2025},
  volume={},
  number={},
  pages={1907-1912},
  doi={10.1109/CDC57313.2025.11312545}}

@article{fire_study_no_tech,
  title={Fighting flames and forging firelines: wildfire suppression effectiveness at the fire edge},
  author={Plucinski, Matt P},
  journal={Current Forestry Reports},
  volume={5},
  number={1},
  pages={1--19},
  year={2019},
  publisher={Springer},
  doi={10.1007/s40725-019-00084-5}
}

@manual{mosek,
   author  = "{MOSEK ApS}",
   title = {MOSEK Fusion API for Python 10.2.16},
   year = {2024},
   url = {https://docs.mosek.com/10.2/pythonfusion/index.html}
 }

@book{mesbahi2010graph,
title = {Graph Theoretic Methods in Multiagent Networks},
author = {Mehran Mesbahi and Magnus Egerstedt},
publisher = {Princeton University Press},
address = {Princeton},
doi = {doi:10.1515/9781400835355},
isbn = {9781400835355},
year = {2010},
lastchecked = {2025-03-09}
}

@misc{distributed_opt_survey,
      title={A Survey of Distributed Optimization Methods for Multi-Robot Systems}, 
      author={Trevor Halsted and Ola Shorinwa and Javier Yu and Mac Schwager},
      year={2021},
      eprint={2103.12840},
      archivePrefix={arXiv},
      primaryClass={cs.RO},
      url={https://arxiv.org/abs/2103.12840}, 
}

@article{fire_detect_large,
  title={Wildfire detection in large-scale environments using force-based control for swarms of UAVs},
  author={Tzoumas, Georgios and Pitonakova, Lenka and Salinas, Lucio and Scales, Charles and Richardson, Thomas and Hauert, Sabine},
  journal={Swarm Intelligence},
  volume={17},
  number={1},
  pages={89--115},
  year={2023},
  publisher={Springer},
  doi={10.1007/s11721-022-00218-9}
}

@InProceedings{fire_supp_large,
title={Extinguishing wildfires in large scale scenarios using swarms of UAVS},
  author={Tzoumas, Georgios and Salina, Lucio and McConville, Alex and Richardson, Tom and Hauert, Sabine},
  booktitle={International Conference on Swarm Intelligence},
  pages={71--83},
  year={2024},
  organization={Springer},
  doi={10.1007/978-3-031-70932-6_6}
}

@article{gan2014online,
  title={Online decentralized information gathering with spatial--temporal constraints},
  author={Gan, Seng Keat and Fitch, Robert and Sukkarieh, Salah},
  journal={Autonomous Robots},
  volume={37},
  number={1},
  pages={1--25},
  year={2014},
  publisher={Springer},
  doi={10.1007/s10514-013-9369-5}
}

@INPROCEEDINGS{8460846,
  author={Viseras, Alberto and Xu, Zhe and Merino, Luis},
  booktitle={2018 IEEE International Conference on Robotics and Automation (ICRA)}, 
  title={Distributed Multi-Robot Cooperation for Information Gathering Under Communication Constraints}, 
  year={2018},
  volume={},
  number={},
  pages={1267-1272},
  doi={10.1109/ICRA.2018.8460846}}

@inproceedings{luo2019distributed,
  title={Distributed environmental modeling and adaptive sampling for multi-robot sensor coverage},
  author={Luo, Wenhao and Nam, Changjoo and Kantor, George and Sycara, Katia},
  booktitle={Proceedings of the 18th International Conference on Autonomous Agents and MultiAgent Systems},
  pages={1488--1496},
  year={2019}
}

@article{NANAVATI2024102503,
title = {Distributed multi-robot source term estimation with coverage control and information theoretic based coordination},
journal = {Information Fusion},
volume = {111},
pages = {102503},
year = {2024},
issn = {1566-2535},
doi = {https://doi.org/10.1016/j.inffus.2024.102503},
url = {https://www.sciencedirect.com/science/article/pii/S1566253524002811},
author = {Rohit V. Nanavati and Matthew J. Coombes and Cunjia Liu}
}

@article{doi:10.1155/2015/286080,
author = {Chengliang Wang and Fei Ma and Junhui Yan and Debraj De and Sajal K. Das},
title ={Efficient Aerial Data Collection with UAV in Large-Scale Wireless Sensor Networks},

journal = {International Journal of Distributed Sensor Networks},
volume = {11},
number = {11},
pages = {286080},
year = {2015},
doi = {10.1155/2015/286080},

URL = { 
    
        https://doi.org/10.1155/2015/286080
    
    

},
eprint = { 
    
        https://doi.org/10.1155/2015/286080
    
    

}
}

@ARTICLE{7365431,
  author={Schwager, Mac and Vitus, Michael P. and Powers, Samantha and Rus, Daniela and Tomlin, Claire J.},
  journal={IEEE Transactions on Control of Network Systems}, 
  title={Robust Adaptive Coverage Control for Robotic Sensor Networks}, 
  year={2017},
  volume={4},
  number={3},
  pages={462-476},
  doi={10.1109/TCNS.2015.2512326}}

@INPROCEEDINGS{7989245,
  author={Kemna, Stephanie and Rogers, John G. and Nieto-Granda, Carlos and Young, Stuart and Sukhatme, Gaurav S.},
  booktitle={2017 IEEE International Conference on Robotics and Automation (ICRA)}, 
  title={Multi-robot coordination through dynamic Voronoi partitioning for informative adaptive sampling in communication-constrained environments}, 
  year={2017},
  volume={},
  number={},
  pages={2124-2130},
  doi={10.1109/ICRA.2017.7989245}}

@INPROCEEDINGS{11128099,
  author={Kailas, Siva and Deolasee, Srujan and Luo, Wenhao and Kim, Woojun and Sycara, Katia},
  booktitle={2025 IEEE International Conference on Robotics and Automation (ICRA)}, 
  title={Integrating Multi-Robot Adaptive Sampling and Informative Path Planning for Spatiotemporal Natural Environment Prediction}, 
  year={2025},
  volume={},
  number={},
  pages={11413-11419},
  doi={10.1109/ICRA55743.2025.11128099}}

@ARTICLE{SIngleDropStudy,
  author={Saikin, Diego A. and Baca, Tomas and Gurtner, Martin and Saska, Martin},
  journal={IEEE Robotics and Automation Letters}, 
  title={Wildfire Fighting by Unmanned Aerial System Exploiting Its Time-Varying Mass}, 
  year={2020},
  volume={5},
  number={2},
  pages={2674-2681},
  doi={10.1109/LRA.2020.2972827}}

@ARTICLE{11005691,
  author={Diaz-Vilor, Carles and Barzegaran, Mohammadreza and Jafarkhani, Hamid},
  journal={IEEE Transactions on Wireless Communications}, 
  title={Multi-UAV Energy-Efficient Wildfire Coverage Optimization}, 
  year={2025},
  volume={24},
  number={10},
  pages={8633-8648},
  doi={10.1109/TWC.2025.3567953}}

@article{fire_impactedbyclimatechange,
title = {Impacts of climate change on the fate of contaminants through extreme weather events},
journal = {Science of The Total Environment},
volume = {909},
pages = {168388},
year = {2024},
issn = {0048-9697},
doi = {https://doi.org/10.1016/j.scitotenv.2023.168388},
url = {https://www.sciencedirect.com/science/article/pii/S004896972307016X},
author = {Shiv Bolan and Lokesh P. Padhye and Tahereh Jasemizad and Muthusamy Govarthanan and N. Karmegam and Hasintha Wijesekara and Dhulmy Amarasiri and Deyi Hou and Pingfan Zhou and Basanta Kumar Biswal and Rajasekhar Balasubramanian and Hailong Wang and Kadambot H.M. Siddique and Jörg Rinklebe and M.B. Kirkham and Nanthi Bolan}
}

@article{fire_impact_onclimate,
author = {John T. Abatzoglou  and A. Park Williams },
title = {Impact of anthropogenic climate change on wildfire across western US forests},
journal = {Proceedings of the National Academy of Sciences},
volume = {113},
number = {42},
pages = {11770-11775},
year = {2016},
doi = {10.1073/pnas.1607171113},
eprint = {https://www.pnas.org/doi/pdf/10.1073/pnas.1607171113}}

@article{fire_by_man,
author = {Jennifer K. Balch  and Bethany A. Bradley  and John T. Abatzoglou  and R. Chelsea Nagy  and Emily J. Fusco  and Adam L. Mahood },
title = {Human-started wildfires expand the fire niche across the United States},
journal = {Proceedings of the National Academy of Sciences},
volume = {114},
number = {11},
pages = {2946-2951},
year = {2017},
doi = {10.1073/pnas.1617394114},
URL = {https://www.pnas.org/doi/abs/10.1073/pnas.1617394114},
eprint = {https://www.pnas.org/doi/pdf/10.1073/pnas.1617394114}}
\end{document}